\documentclass[prc,aps,groupedaddress,superscriptaddress,twocolumn,nofootinbib,showpacs,floatfix,a4paper]{revtex4-1}

\usepackage{graphicx,colordvi,rotating}
\usepackage{multirow,color}
\usepackage{amsmath,amssymb}
\usepackage[mathscr]{euscript}
\usepackage{cancel}
\usepackage{braket}
\usepackage{cases}
\usepackage{color}
\usepackage{mathtools}
\usepackage{natbib}
\usepackage{ulem}

\begin{document}

\title{Description of the asymmetric to symmetric fission transition in the neutron-deficient Thoriums - Role of the tensor force }

\author{R. N. Bernard}
\altaffiliation{Corresponding author: remi.bernard@anu.edu.au}
\affiliation{Department of Nuclear Physics, Research School of Physics and Engineering, Australian National University, Canberra, Australian Capital Territory 2601, Australia}
\affiliation{CEA, DAM, DIF, F-91297 Arpajon, France.}
\affiliation{Ecole normale sup\'erieure Paris-Saclay, 61 Avenue du Pr\'esident Wilson, 94230 Cachan, France}

\author{N. Pillet}
\altaffiliation{Corresponding author: nathalie.pillet@cea.fr}
\affiliation{CEA, DAM, DIF, F-91297 Arpajon, France.}

\author{L. M. Robledo}
\affiliation{Departamento de F\'{\i}sica Teorica, Universidad Aut\'onoma de Madrid, E-28049 Madrid, Spain.}

\author{M. Anguiano}
\affiliation{Departamento de F\'{\i}sica At\'omica, Molecular y Nuclear, Universidad de Granada, E-18071, Granada, Spain.}

\date{\today}

\def\dspt{\displaystyle}

\begin{abstract}
In the present study, we have investigated the impact of the tensor 
force on fission paths, in particular the symmetric and asymmetric 
barriers in $^{230}$Th, $^{226}$Th, $^{222}$Th and $^{216}$Th isotopes 
which display an asymmetric to symmetric fission transition. This 
analysis has been performed within the HFB approach with 
(Q$_{20}$,Q$_{30}$,Q$_{40}$) as collective variable constraints, using 
the D1ST2a Gogny+tensor term interaction and comparing to the standard 
D1S Gogny interaction results. The effects from the tensor term on the 
potential energy surface landscape, and especially on barrier heights 
and its topology by opening a new valley in agreement with experimental 
data, are found to be crucial in the description of exotic actinide 
fission. We conclude that a tensor term should be integrated to the 
long range part of the effective interaction for a better description 
of the fission.
\end{abstract}


\maketitle

\section{Introduction}

The fission of the nucleus into two or more fragments is one of the 
most complex phenomenon in nuclear physics. Its complete modeling 
requires the knowledge of both static and dynamic properties of the 
fissioning system, namely the static nuclear configurations out of 
equilibrium, the coupling between collective and intrinsic degrees of 
freedom and the dynamics of large amplitude collective motion. Along 
the years, different types of approaches and models have been developed 
to tackle this difficult problem. Among them, fully microscopic 
approaches allow a description of the entire process from the initial 
configuration up to the scission point and beyond 
\cite{schunck,walid1,Gogny4,Gogny5,Goutte,Dubray,walid2,walid3,GiuPLB18,BerPRC11,warda,%
Regnier1,walid4,zdeb,walid5,Regnier2,scamps,lacroix1,lacroix2,bulgac}. 
Even though this very ambitious program is far from being completed to date, 
it offers the possibility to take into account, in a unified and 
coherent way, both collective and internal degrees of freedom (and its 
interaction) along the fission path within a fully quantum-mechanical 
description of the time-dependent evolution of the fissioning nucleus. 
These approaches are based on a mean field description and therefore 
have to rely on the properties of effective nucleon-nucleon 
interactions, whose parameters are the only inputs of the model. Those 
parameters are fixed {\it a priori} once and for all using some fitting 
protocol which may or may not include fission data for a relevant set 
of nuclei.

In the past, most of the progress made in the field of the microscopic 
description of fission was accomplished essentially through a better 
understanding of the nuclear effective force. The first completely 
microscopic calculation was performed by the Orsay group with an early 
parametrization of the Skyrme interaction \cite{Flocard1,Flocard2}. 
They calculated the symmetric fission barrier in $^{240}$Pu within the 
HF+BCS approach under a constraint on the mass quadrupole moment. Even 
though the structure of the double-hump fission path was correctly 
described, the heights of the barriers were too high as compared to the 
experimental data extracted from measurements of neutron induced cross-sections. This 
was a common feature of many microscopic calculations carried out with 
different Skyrme parameterizations and for various actinides 
\cite{Negele,Brack}. The analysis made by Dutta et {\it al.} for 
several Skyrme forces led to the conclusion that the second barrier 
height scaled like the value of the surface coefficient $a_{s}$ of the 
interactions \cite{Dutta}. Similar studies were done with the original 
D1 Gogny force \cite{Gogny4,Gogny5,Gogny1,Gogny2,Gogny3,Gogny9}. 
The results displayed similar features as the ones obtained with the 
Skyrme interactions, at very large deformations. In order to improve
the agreement with experiment for the fission barriers of 
the typical benchmark nucleus $^{240}$Pu, the surface 
tension coefficient of the D1 force was decreased leading to the 
well-known D1S parameterization \cite{Gogny4,Gogny5,Gogny6}. 
Most of the properties of the D1S parameterization are similar to the ones of D1 
but the barrier heights are in general in a much better agreement with 
the experimental ones. Another consequence of the fit was a desired weaker pairing strength in D1S
with the corresponding impact on the collective inertias \cite{girod}. 
This is a direct consequence of the dependence of the inertias with the inverse
of the pairing gap \cite{Bra72,Ber91,Mor74}.
Once the static deformation  and pairing properties of the force were fixed, 
it was conceivable to
think on how to improve the treatment of the dynamics of the fission process. A 
dynamical treatment using the Time-Dependent Generator Coordinate 
Method plus the Gaussian Overlap Approximation with the Gogny 
interaction was proposed in the eigthies \cite{Gogny4,Gogny5,Gogny6} with additional 
refinements thereafter \cite{Goutte,Dubray,BerPRC11, Regnier1,Regnier2,GiuPLB18,Giu14,Ber19}. 
The Gogny force in its various incarnations has been used not only in fission,
but also in the rather successful description of many low energy nuclear properties
at the mean field and beyond (see Refs. \cite{RobJPG19} and \cite{Peru} for recent reviews).

In the early versions of the Skyrme and Gogny effective forces the tensor force, similar to the
one in the one pion exchange potential, was disregarded in order to simplify calculations. 
Recently, this term has received renewed attention in
connection with properties of both spherical and deformed nuclei described with the Skyrme interaction 
\cite{lesinski1,lesinski2}.        
In the case of the Gogny interaction, a few attempts tackled this issue. One cites
the pioneer work of \cite{Otsuka} where only the like-particle component of the tensor 
force was included. The aim of the introduction of the tensor term was 
to improve the evolution of spherical single-particle states along isotopic chains. 
A full refitting of the Gogny force was carried out. However, no attention was paid to the pairing properties.
The perturbative addition of a complete, long range tensor 
term proposed by M. Anguiano et {\it al.} \cite{marta1,marta2,marta3,marta4,marta5} is a 
fully meaningful work in the case of the Gogny force. Indeed, the Gogny force was partly adjusted on the 
results obtained from a G-matrix plus second order corrections 
\cite{maire,gognypires}, leaving room for reasonable extensions of the 
mean-field to treat explicitly the nuclear long range correlations \cite{chappert,Gogny7}. 
The main result obtained by M. Maire and D. Gogny, using the soft and local GPT effective force, was that the 
second order corrections coming from the tensor force mostly affected the (S=1,T=0) channel.
In the construction of the effective Gogny force, most of the effect of the tensor force was taken into account in 
the strength of the density-dependent central term which also acts in 
the (S=1,T=0) channel. Thus, the parameters of the standard Gogny 
interaction already take into account in a phenomenological way most of 
the effect of the tensor and, as a consequence, only a residual tensor 
with a long range is needed to fully take into account the effect of 
this part of the nuclear force. This results strongly softens the 
conclusions reached in the context of Skyrme interactions concerning 
the necessity to fully readjust the parameters of the interaction and 
the inadequacy of  a perturbative addition of a tensor term.
Of course, a complete refitting of all the 
parameters of the Gogny force would be highly desirable and this is an 
objective to be pursued in the short term. However, the present perturbative tensor
allows one to look for new experimental data sensitive to the 
physics of a residual tensor term in order to constrain the additional 
parameters introduced. The study presented in this paper has been done 
in the same spirit.

To our knowledge, the impact of the tensor term in the potential energy 
surface and collective inertia required for fission has never been 
investigated. On the other hand, the role played by the tensor term has been investigated recently in 
several fusion studies \cite{Sekizawa,guo,umar1,umar2,stevenson}.
The tensor interaction rearranges the position of the single particle 
orbitals changing the shell effects responsible for many of the properties of
the quantities relevant to fission. In fact, it could be the missing ingredient required 
to explain a symmetric bimodal fission mode recently found in some 
neutron-deficient thorium isotopes \cite{Martin,Pellereau,SOFIA1,SOFIA2}. 
Recent experimental data provided by the experiments of the SOFIA collaboration
have revealed the existence of such a symmetric 
bimodal mode, composed of the standard super-long mode and a new 
compact mode. This latter is characterized by a non-ambiguous decrease 
of the mean value of the total prompt neutron multiplicity along the asymmetric to symmetric 
fission transition in the neutron-deficient thorium isotopes. In the 
present article, we discussed the role of the tensor interaction in the 
context of the Gogny force for the description of this new compact 
mode, using a Hartree-Fock-Bogoliubov approach (HFB) with several 
constraints. In particular, the role as collective variables of the 
Q$_{20}$, Q$_{30}$ and Q$_{40}$ axially symmetric multipole moments is 
investigated. 

The article is organized as follows. In section \ref{model}, the 
ingredients of the model used in the present study are discussed: the 
HFB method with constraints is briefly summarized and the D1ST2a 
parameterization of the Gogny interaction is described. This 
parameterization is an extension of D1S in which a perturbative tensor 
term is added. In section \ref{resultsA}, symmetric and asymmetric 
fission paths are shown for the $^{216-232}$Th isotopes. In this first 
analysis, the Q$_{20}$ and Q$_{30}$ collective variables are 
considered. The impact of the tensor term on the first and second 
barrier heights is discussed. In section \ref{resultsB}, the role of 
the Q$_{40}$ multipole moment is highlighted and an explanation of the 
origin of the new compact symmetric fission mode is proposed. In 
section \ref{resultsC}, the various contributions of the D1S and D1ST2a 
interactions to the HFB binding energies are detailed. In section 
\ref{resultsD}, the distribution of the available energy at scission is 
discussed and evaluated in order to obtain general trends concerning 
the number of emitted neutrons in the case of the super-long and the 
compact mode. Finally, in section \ref{conclusion}, conclusions and 
perspectives are given.

\section{Static microscopic model}\label{model}

\subsection{Hartree-Fock-Bogoliubov method with constraints}

As it is widely recognized, the mean-field and its extensions are powerful 
approaches to  describe the wave function of the ground and excited states of 
the nucleus. On the other hand, fission is a time dependent phenomenon 
and a couple of extensions to the traditional stationary mean field are 
of use in its study: one is the Time-Dependent Hartree-Fock-Bogoliubov (TDHFB) method 
which is the standard time dependent generalization of the 
Hartree-Fock-Bogoliubov method and the other is the Time-Dependent 
Generator Coordinate Method (TDGCM) which is a fully quantum mechanics 
procedure. In both cases, the determination of the potential energy surface (PES),
i.e. the HFB energy as a function of several relevant constraints is essential
in determining the dynamics of the system and a lot of information can be
gained by studying its evolution with the relevant degrees of freedom.
In this paper, we have restricted ourselves to the study of the PES as a
function of axially symmetric multipole variables in order to understand
the impact of the tensor term. A full dynamical study in the framework
of the TDHFB \cite{SimenelPPNP} or TDGCM will be the subject of future studies.
Both approaches have their
own advantages and drawbacks. The TDGCM, which has been developed
within the Gaussian Overlap Approximation (GOA), contains two main steps
\cite{walid1}
\begin{enumerate}
\item a static calculation which determines the PESs and collective inertia, 
      using the HFB method under constraints on relevant collective variables. 
      The only ingredient is the nuclear effective interaction.
\item a dynamic calculation in collective space and based on the previously 
      determined input which describes the time evolution of the system 
      up to the scission.
\end{enumerate} 
With this method one can obtain, for example, the fission fragment yield 
distributions. However, there is still room for improvement within
the TDGCM+GOA framework, and one can mention the following improvements 
\begin{itemize}
\item Remove some of the approximations used to compute the inertias, eventually
      using the exact ones \cite{GiuPLB18}
\item Include intrinsic excitation \cite{walid1,BerPRC11} to describe dissipation
\item Restoration of broken symmetries, like angular momentum or particle number \cite{Ber19}
\item Removal of the GOA
\item Exploration of alternative effective interactions.
\end{itemize}
In the present work, we will pursue the issue of studying additional terms in the interaction,
and for simplicity, we will restrict ourselves to the static part of the calculation. With this
in mind we have analyzed the influence of the tensor term on the PES topology.

The HFB equation has been solved by conserving the axial, time-reversal and simplex symmetries. The parity
has been broken in order to study the asymmetric fission through 
non-zero odd multipole moment paths. Moreover, two types of constraints
have been considered: the first one concerns the average conservation
of proton and neutron numbers, the second one is dedicated to multipole
moments. Thus, the minimization principle on the total energy of the
system reads:
\begin{equation}
 \delta \bra{\Phi}   \hat H - \lambda_n \hat Q_n - \lambda_p \hat Q_p - \sum_{i} \lambda_i \hat Q_{i0} \ket{\Phi} = 0
\end{equation}
where $\hat H$ is the nuclear Hamiltonian, $\hat Q_n$ and $\hat Q_p$ the particle number operators 
and $\hat Q_{i0}$ the multipole moment operators defined as:
\begin{equation}
 \hat Q_{i0} = \sqrt{\frac{4\pi}{2i+1}}\sum_{l=1}^{A}r_l^i Y_{i0}(\theta_l,\phi_l)
\end{equation}
The set of \{$\lambda_{i}$\} are the Lagrange parameters associated 
with the corresponding constraint operators. The $i$th order multipole 
moment variable $q_{i0}$ is defined as the average of the $i$th order 
multipole operator $Q_{i0}$ in the HFB state $\ket{\Phi}$. The monopole 
moment $q_{10}$ is set to zero in order to avoid contamination with 
spurious solutions coming from the breaking of the spatial translation 
symmetry. In the following, the one-dimensional (1D) fission paths are 
calculated using only the quadrupole moment variable as collective 
degree of freedom in addition to $q_{10}$. The symmetric path will 
refer to HFB calculations where $q_{i0}$ are set to $0$ fm$^{i}$ for 
odd $i$, $i>1$ whereas these latter are let free for the asymmetric 
path. Besides, the two-dimensional (2D) fission PES are obtained with 
two constrained multipole moments, for example \{Q$_{20}$, Q$_{30}$\} 
or \{Q$_{20}$, Q$_{40}$\}. As for the 1D path, when the symmetric 
fission is studied, the $q_{i0}$ are set to $0$ fm$^{i}$ for odd $i$.

We have implemented the tensor term in a computer code, named HFBaxialT 
\cite{HFBaxialT}, which is built upon the HFBaxial code 
\cite{HFBaxial}, and uses an expansion of the quasiparticle operators 
in a harmonic oscillator basis to solve the HFB equation. In the 
HFBaxialT code the approximate second order gradient method is employed 
to minimize the HFB energy \cite{Ber.11}. The main advantage of this 
over other traditional iterative methods is the easy handling of 
constraints and an almost perfect rate of sucess in reaching a 
converged HFB solution.

The quasiparticle operators are expanded in an axially symmetric harmonic oscillator basis with
a maximum value of quanta in the perpendicular direction $N_\perp=2n_\perp + |m|$ of 14 and a maximum value of quanta in the
$z$ direction $n_z$ of 21. Although the basis size is rather limited for the calculation of absolute values, it is enough
for the calculation of relative effects, like energy differences (see below). The two oscillator lengths of the basis
$b_\perp$ and $b_z$ have been optimized as to minimize the HFB energy for each constrained calculation.
The HFB solutions are computed from sphericity up to 
scission within a mesh defined by the step-size $2$ b, $4$ b$^{3/2}$ and $5$ b$^2$ along the quadrupole, octupole and hexadecapole moment variables, respectively. 

\subsection{The D1ST2a Gogny+tensor interaction}

The present study has been done in the context of the Gogny interaction. As discussed in the introduction, the HFB 
mean-field obtained from the D1S Gogny interaction, which is historically known as the reference Gogny interaction to
performed fission studies, takes into account in a phenomenological way most of the effect of the tensor term through
its (S=1,T=0) zero-range, density-dependent central term component.
However, the effect of the long range part of a residual tensor is still missing and expected to play a role
in specific situations, as for example an accurate description of the spin-orbit splittings, the un-natural parity states, the 
proton-neutron pairing or deformation properties.

The D1ST2a interaction is characterized by the adding of a perturbative tensor with long range
and a weak strength to the Gogny D1S interaction \cite{marta1,marta2,marta3,marta4,marta5}. Its analytical form reads as:
\begin{equation}
\begin{array}{l}
 V^{\rm D1TS2a}(\vec{r}) = \\
\displaystyle \hspace{0.5cm} \sum_{i=1}^{2} (W_{i} + B_{i}P_{12}^\sigma + H_{i}P_{12}^\tau + M_{i}P_{12}^\sigma P_{12}^\tau) e^{-\vec{r}^2/\mu_{i}^2 } \\
\hspace{0.4cm} + t_0 (1+x_0P_{12}^\sigma)  \rho^\alpha(\vec{r}) \delta(\vec{r})  \\
\hspace{0.4cm} + W_{LS} \overleftarrow{\nabla} \delta(\vec{r}) \land \overrightarrow{\nabla} . (\vec{\sigma_1}.\vec{\sigma_2})  \\
\hspace{0.4cm} +  ( V_{\rm T1}+V_{\rm T2}\,  P_{12}^\tau ) \hat S_{12}(\vec{r}) e^{-\vec{r}^2/\mu_{\rm TS}^2 }
\end{array}
\end{equation}
where the first three components correspond to the Gogny interaction with the D1S parameterization. In the above expression, the 
$P_{12}^\sigma$ and $P_{12}^\tau$ operators are the traditional spin and isospin exchange operators, respectively. The set of parameters 
\{W$_{i}$, B$_{i}$, H$_{i}$, M$_{i}$, $i=1,2$ \}, t$_{0}$ and W$_{\rm LS}$ are the coefficients of central, density-dependent central and 
spin-orbit terms. The \{ $\mu_{i}$, $i=1,2$ \} are the ranges of the Gaussian form factor and the coefficient x$_{0}$ is set to one to 
prevent the contribution of the density-dependent term to the proton and neutron pairing channels. Finally, $\vec \sigma$ is the 3-dimension 
spin operator and the operator $\hat S_{12} (\vec{r})$ is the usual tensor operator 
which is defined as:
\begin{equation}
\displaystyle \hat S_{12} (\vec{r}) \, = \, 3 \, \,  \frac{\,  \vec{\sigma_1} \cdot \vec{r} \,\,  \vec{\sigma_2} \cdot \vec{r} \, }{r^{2}} \, - \, \vec{\sigma_1} \cdot \vec{\sigma_2}
\end{equation}

The parameters of the non-tensor terms of D1ST2a are the same as the 
D1S parameters. On the other hand, the parameters $V_{\rm T1}$ and 
$V_{\rm T2}$  of the tensor term are adjusted to reproduce the neutron 
single particle energies $1f_{5/2}$ and $1f_{7/2}$ in ${}^{48}$Ca. The 
range $\mu_{\rm TS}=\mu_{2}=1.2$ fm in the Gaussian form factor of the 
tensor has been chosen equal to the longest range of the two Gaussians 
in the central potential. We consider only the contribution of the 
tensor term to the mean field part of the HFB method whereas its 
contribution to the pairing channel is not taken into account. 
Therefore, and as in the D1S case, the only term contributing to the 
pairing channel in D1ST2a is the central potential. The reason for this 
omission is that the residual tensor term is not expected to play a 
relevant role in the proton and neutron pairings, unlike the 
proton-neutron one. Besides, the Coulomb exchange term of the HFB 
Hamiltonian is computed with the Slater approximation in the two cases.
 
Since in this study we are mainly interested in the impact of the 
tensor term in the fission process, special attention is paid to the 
total binding energy difference between the results obtained with the 
D1S and the D1ST2a interactions. In order to justify our choice of 
harmonic oscillator basis size, we have checked $\Delta(\rm Nsh)=E^{\rm 
Nsh}_{\rm HFB}({\rm D1S})- E^{\rm Nsh}_{\rm HFB}({\rm D1ST2a})$ is the 
same for N$_{\rm sh}$=14 and N$_{\rm sh}$=15. Here $N_{\rm sh}$ is the 
maximum number of quanta in the perpendicular direction and it is 
consider as the equivalent of the number of shells in an spherical 
basis. The maximum values of $n_z$ in the two cases are 21 and 23, 
respectively. The energy difference 
\begin{equation}
 \Delta {\rm E} = \big| \Delta(15)-\Delta(14) \big|
\end{equation}
has been calculated along the symmetric path of $^{226}$Th. It 
averages $23$~keV along the whole fission path and reaches $104$~keV 
at large deformation which is still negligible compared to energy 
differences between the D1S and D1ST2a fission paths presented below.

The interplay between the D1ST2a tensor term and the quadrupole 
deformation properties was recently studied on various isotopic chains 
especially in the $sd$-shell \cite{marta5}. However, typical  
quadrupole deformation values are much lower than the extreme ones 
encountered during the fission process. The results of this first study 
lead to the conclusion that, depending on the filling of the shells, 
the tensor term may strongly influence the HFB total energy, modify the 
potential energy landscape and change the ground state deformation. 
Pairing properties are also affected, especially a weakening of the 
particle number fluctuations is observed. An interpretation of such an 
influence in terms of spin-isospin contributions to the HFB energy has 
been given: most of the time, the tensor term give rise to a repulsive 
dominant proton-neutron contribution to the HFB energy. Attractive 
like-particle contribution become dominant when the filling of the 
valence shells are in a Spin-Saturated/Spin-Unsaturated configuration, 
which happens around sphericity. As a logical continuation of this 
latter study, the present investigation is of prime interest as fission 
properties are strongly sensitive to PES landscapes in terms of 
collective variables and pairing degree of freedom.


\section{Results and discussion}\label{results}

There are many observable which are required to fully characterize and 
understand fission. One can cite for example the mass and charge 
distribution of the fragments, the total kinetic energy (TKE) and the 
average neutron multiplicities $\langle \nu \rangle$.

The pioneer experiments of K.H. Schmidt at GSI Darmstadt  based on the 
production of an exotic secondary beam by fragmentation of a primary 
beam of $^{238}$U at relativistic energies, followed by Coulomb 
excitation of the secondary beam \cite{Schmidt} opened up the door to 
the measurement of the charge distribution of the fragments in 
neutron-deficient actinides and preactinides. The isotopes 
$^{205,206}$At, $^{204-209}$Rn, $^{206-212,217,218}$Fr, $^{209-219}$Ra, 
$^{212-226}$Ac, $^{217-229}$Th, $^{224-232}$Pa and $^{230-234}$U were 
considered in a series of experiments. The results pointed out to a 
transition from  asymmetric to symmetric fission in this region of the 
nuclear chart. However, the TKE measured in the $^{210-215,217-219}$Ra, 
$^{215-223}$Ac, $^{221-229}$Th, $^{226-232}$Pa and $^{232-234}$U 
isotopes were known with low accuracy. Both the masses of the fragments 
and the average neutron multiplicity were not accessible in this kind 
of experiments. Starting from the same reaction mechanism but using a 
much more advanced experimental setup, the SOFIA experiments at GSI 
Darmstadt now allow one to obtain both charge and mass fission yields with 
an accuracy smaller than a mass unit as well as the average neutron 
multiplicity $\langle \nu \rangle$ \cite{Martin,Pellereau}. Other 
techniques have been developed in parallel to study the fission of 
exotic nuclei. An example is the $\beta$-delayed fission process used 
at ISOLDE to study the fission of the very exotic nucleus $^{180}$Hg 
which surprisingly shows asymmetric fission \cite{Andreyev}. Another 
example are the transfer and fusion reactions using a beam of $^{238}$U 
at $6$~MeV per nucleon on a $^{12}$C target used in the  GANIL laboratory 
along with the VAMOS spectrometer \cite{CaaPLB17}.

The study of the fission of the neutron-deficient Thorium isotopes, 
which are analyzed in this paper, has been motivated by the 
experimental data obtained during the 2012 SOFIA campaign at GSI 
Darmstadt \cite{SOFIA1,SOFIA2}. The measurements concern the thorium 
isotopes $^{230}$Th, $^{229}$Th, $^{226}$Th, $^{225}$Th, $^{223}$Th, 
$^{222}$Th and $^{221}$Th. The experimental results confirm the 
asymmetric to symmetric transition in the mass distribution of the 
fragments already observed in Ref \cite{Schmidt}. In addition, they 
suggest the existence of a new bimodal symmetric fission mode in this 
region, composed of the standard super-long mode plus a new compact 
one. The compact component is experimentally characterized by the 
strong decrease of the average neutron multiplicity along the isotopic 
chain for decreasing neutron number. 

\subsection{Symmetric and asymmetric fission paths using Q$_{20}$ and Q$_{30}$ as collective variables}\label{resultsA}

In this section, we analyze two fission paths, the symmetric and the 
asymmetric one, the latter being obtained by minimizing the total HFB 
energy for a non-zero average value of Q$_{30}$. In a first step, the 
global 1D axial deformation properties of the even-even $^{216-232}$Th 
isotopes are discussed. In a second step, the \{Q$_{20}$, Q$_{30}$\} 
PES's are analyzed for the $^{216}$Th, $^{222}$Th, $^{226}$Th and 
$^{230}$Th isotopes.

\subsubsection{Global axial deformation properties of the even-even $^{216-232}$Th}

\begin{figure}[htb] \centering
\begin{tabular}{cc}
   \includegraphics[width=8.5cm]{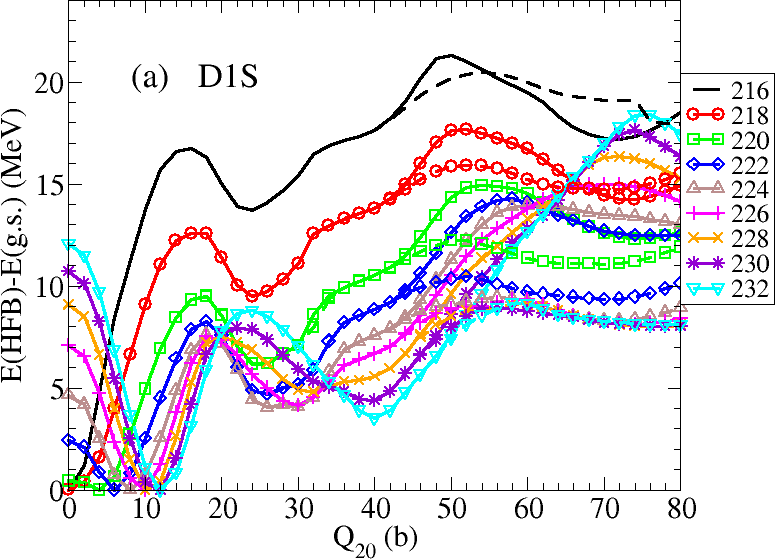} \\  \vspace{0.5cm}
     \includegraphics[width=8.5cm]{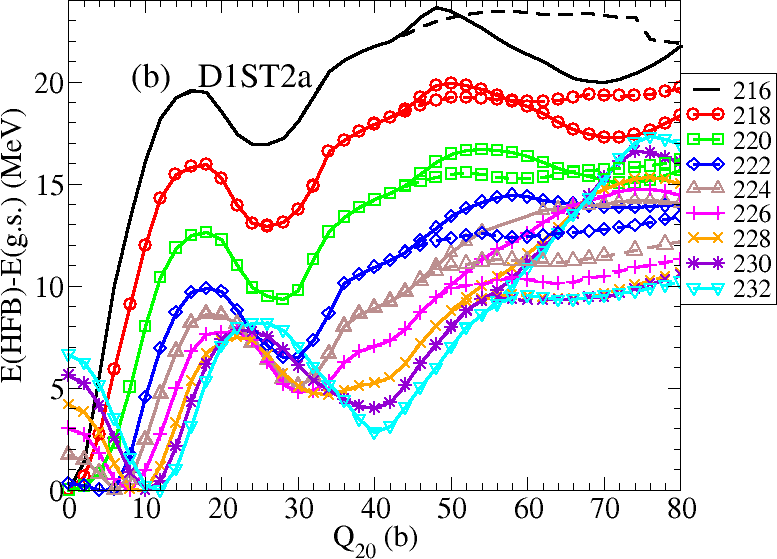}
\end{tabular}
\caption{Barrier heights of $^{216}$ up to $^{232}$Th even-even isotopes for the symmetric (full lines) and asymmetric (dashed lines) paths, 
calculated at the HFB level with the D1S (top) and D1ST2a (bottom) Gogny interactions. Energies are expressed in MeV.} 
\label{Ap1}
\end{figure}

The evolution of the HFB total energy calculated with the D1S (D1ST2a) 
Gogny force is shown up to the second barrier in Fig. \ref{Ap1} (a) 
(Fig .\ref{Ap1} (b))  for even-even Thorium isotopes with the mass A 
ranging from 216 to 232. The symmetric paths are represented by the 
full lines and the asymmetric ones by the dashed lines. In 
order to facilitate the comparison the HFB ground state binding energy 
has been subtracted for each isotopes. 

Concerning the symmetric path, one observes large variations in the 
position of the minima and the maxima of the potential energy curves 
(PEC) and the barrier heights. By looking  at the results obtained with 
the D1S Gogny force for the symmetric path (see Fig. \ref{Ap1} (a)), 
one sees that the heaviest isotope, namely $^{232}$Th, is a 
well-deformed nucleus in its HFB ground state is characterized by 
Q$_{20} \simeq 12$~b an energy gain of $\sim 12$~MeV with respect to 
the spherical configuration. The maximum of the first hump is obtained 
at  Q$_{20} \simeq 24$~b and its height is $\sim 9$~MeV. The second 
well (fission isomer) is located around Q$_{20} \simeq 40$~b and is 
$3.5$~MeV higher in energy than the HFB ground state. At Q$_{20} \simeq 
74$~b, one encounters the second hump whose height is $\sim 18$ ~MeV. 

By exploring the isotopic chain from $^{232}$Th to $^{216}$Th, one 
observes that the first well is continuously less and less deformed, 
the deformation energy decrease, in such a way that in $^{218}$Th and 
$^{216}$Th the HFB ground state is spherical. This effect is driven by 
the N=126 neutron shell which corresponds to the $^{216}$Th isotope. 
New experimental data have shown 
this shell closure effect at N=126 in Po, Rn, Ra and Th isotopes, with 
a relaxation in the U ones \cite{refsup1,refsup2}. Calculations of the 
excitation energy of the 2$^{+}_{1}$ states (the first excited state in this nuclei) 
using the 5-dimensionnal collective lead to the same conclusion, even though the
relaxation effect is not so visible for U isotopes \cite{refsup3,refsup4}. 
The clear increase of the energy of the 2$^{+}_{1}$ states signs this shell effect,
the increase of the rigidity of the nuclei at N=126, which is accompanied by 
an absence of neutron pairing energy at the minima of the potential energy surfaces.
The heights of the first hump slightly decrease from $^{232}$Th to 
$^{224}$Th and increase again from $^{222}$Th to $^{216}$Th in such a 
way that it reaches $\sim 17$~MeV in the $^{216}$Th isotope. The 
lightest isotopes are predicted to be more rigid than the other ones.

Concerning the second wells, one obtains the same trend as the one 
observed for the ground state wells. The decrease of the associated 
Q$_{20}$ value results in two deformation regions typical of the 
$^{216}$Th isotope around $25$~b and the $^{232}$Th isotope around 
$40$~b. The associated excitation energies are roughly the same from 
$^{230}$Th to $^{222}$Th and start to increase in a significant way in 
$^{220}$Th, $^{218}$Th and $^{216}$Th isotopes. 

Finally, for the second hump, one notes that its quadrupole deformation 
is essentially distributed around two deformation regimes, namely 
Q$_{20} \simeq 70$~b for $^{232}$Th up to $^{226}$Th isotopes, and 
Q$_{20} \simeq 50$~b for the lighter ones. The effect of the N=126 
neutron magic number seems to manifest again at these deformations. The 
heights of the second hump decrease from $^{232}$Th to $^{224}$Th 
isotopes, then start to increase up to the $^{216}$Th one for which it 
is equal to $22$~MeV. Most of the isotopes seems to display a third 
symmetric hump, as it was observed experimentally in the heavier 
isotopes $^{232}$Th, $^{231}$Th and $^{230}$Th 
\cite{blons1,blons2,blons3,theseblons,romain}.

From Fig. \ref{Ap1} (b), one observes that the general trends obtained 
with the D1S Gogny interaction are still valid with the D1ST2a 
interaction. However, the ground state deformation energies are found 
to be in general  much smaller when the tensor term is taken into 
account. The main consequence is that a spherical HFB ground state is 
already obtained for $^{220}$Th. Also, the tensor term is able to 
modify in a non-negligible way the height of the first and second 
humps, by increasing or decreasing them by several MeV depending on the 
isotopes. This last point, already visible by comparing Fig. \ref{Ap1} 
(a) and (b), will be discussed further down.

Concerning the asymmetric path, one notes the following features with 
the D1S Gogny interaction. In the $^{232}$Th isotope, the asymmetric 
path starts to be favorable in energy around Q$_{20} \simeq 50 $~b, 
after the bottom of the second well. The height of the second hump 
which corresponds to the asymmetric path is equal to $\simeq 9.3$~MeV. 
This value has to be compared to the symmetric one which is $\simeq 
18.8$~MeV. As a general rule, the height of the asymmetric second hump 
is always lower in energy than the one of the symmetric second hump. 
Along the isotopic chain, one observes that the opening of the 
asymmetric valley occurs earlier and earlier in deformation and 
stabilizes in the $^{218}$Th and $^{216}$Th isotopes around Q$_{20} 
\simeq 42$~b. Moreover, the difference between the heights of the 
symmetric and the asymmetric second humps decreases continuously and 
regularly, when going towards the lightest isotopes. In the $^{216}$Th 
isotope, it is equal to $\simeq 1.5$~MeV. For comparison, in the 
transitional nucleus $^{222}$Th, it reaches $\simeq 4$~MeV.

With the D1ST2a Gogny interaction, the same observations can be done 
concerning the opening of the asymmetric valley. One notes the increase 
of the first and second humps in the lightest isotopes. The main 
difference comes from the effect of the tensor on the relative position 
of the maxima of the symmetric and asymmetric second humps, which is 
reduced considerably. For comparison with the D1S interaction, it is 
equal to $\simeq 7.5$~MeV in the $^{232}$Th isotope and $\simeq 
1.5$~MeV in the $^{222}$Th isotope. In the $^{216}$Th isotope, this 
difference tends to zero. Then, one concludes that the tensor term of 
the D1ST2a interaction tend to equate  the heights of the symmetric and 
asymmetric second humps, rendering the symmetric path energetically 
competitive in the lightest Thorium isotopes.

We turn now our attention to the various contributions to the total HFB energy, 
namely the mean-field without the tensor (E$_{ \rm MF}$), the pairing (E$_{\rm pair}$) and 
the tensor (E$_{\rm TS}$) ones. In view of that, we have defined the three quantities 
$\Delta {\rm E}_{{\rm HFB}}$, $\Delta {\rm E}_{{\rm MF}}$ and $\Delta 
{\rm E}_{{\rm pair}}$ which represent the difference between the total 
HFB energies, the mean-field energies, pairing energies, respectively, 
calculated with the D1ST2a and the D1S interactions.
\begin{eqnarray}
 \Delta {\rm E}_{{\rm HFB}}  & = &  {\rm E}_{{\rm HFB}}^{{\rm D1ST2a}} - {\rm E}_{{\rm HFB}}^{{\rm D1S}} \nonumber \\
 \Delta {\rm E}_{{\rm MF}}   & = &  {\rm E}_{{\rm MF}}^{{\rm D1ST2a}} - {\rm E}_{{\rm MF}}^{{\rm D1S}}   \nonumber \\
 \Delta {\rm E}_{{\rm pair}} & = &  {\rm E}_{{\rm pair}}^{{\rm D1ST2a}} - {\rm E}_{{\rm pair}}^{{\rm D1S}}   
\end{eqnarray}
In addition we have also considered ${\rm E}_{{\rm TS}}$ which is the tensor contribution obtained with the D1ST2a interaction.
Results are shown in Tables \ref{diff_sph}, \ref{diff_gs}, \ref{diff_bar1} and \ref{diff_swell}  for the spherical and the ground state, the first hump
and the second well of $^{216}$Th, $^{222}$Th, $^{226}$Th and $^{230}$Th, respectively. Values are also given for standard actinides, 
namely $^{236}$U and $^{240}$Pu. All the Thorium isotopes and $^{240}$Pu shells are unsaturated at the Fermi levels. The case of $^{236}$U is 
different as the proton $1h$ valence shell is spin saturated.

\begin{table}[!] \centering
 \begin{tabular}{|c|c|c|c|c|c|c|}
\hline	
     Spher.          &  $^{216}$Th & $^{222}$Th        & $^{226}$Th        & $^{230}$Th        & $^{236}$U         & $^{240}$Pu  \\
\hline
 $\Delta {\rm E}_{{\rm HFB}}$        &  $4.310$   &     $3.713$       &    $3.279$        &    $2.881$        &     $1.557$       &  $1.922$   \\
 $\Delta {\rm E}_{{\rm MF}}$   & $-6.286$   &   $-6.213$     &      $-6.311$     &      $-6.433$     &       $-11.230$   &   $-6.517$   \\
 $\Delta {\rm E}_{{\rm pair}}$ &  $8.056$   &    $7.564$      &      $7.393$      &      $7.310$      &       $12.780$    &   $7.396$   \\
 ${\rm E}_{{\rm TS}}$          &  $2.541$   &    $2.362$        &       $2.198$    &      $2.003$     &     $0.007$       &  $1.043$  \\
\hline 
 \end{tabular}
\caption{Energy differences $\Delta E = E^{{\rm D1ST2a}}-E^{{\rm D1S}}$ in MeV for HFB, mean field and pairing energies. The tensor contribution to the 
D1ST2a HFB energy is also given.}
\label{diff_sph} 
\end{table}

\begin{table}[!] \centering
 \begin{tabular}{|c|c|c|c|c|c|c|}
\hline	
     g.s.          &  $^{216}$Th & $^{222}$Th        & $^{226}$Th        & $^{230}$Th        & $^{236}$U         & $^{240}$Pu  \\
\hline
 $\Delta {\rm E}_{{\rm HFB}}$     & $4.310$     &       $5.843$     &      $7.336$     &      $7.968$      &       $8.667$    &   $8.782$  \\
 $\Delta {\rm E}_{{\rm MF}}$  & $-6.286$    &  $4.336$          &  $8.337$         &  $6.359$          &   $5.576$        &   $4.755$   \\
 $\Delta {\rm E}_{{\rm pair}}$ &  $8.056$    &    $-1.941$       &    $-5.847$      &    $-3.947$       &    $-3.629$      &  $-3.016$    \\
 ${\rm E}_{{\rm TS}}$        & $2.541$     &  $3.448$          &    $4.846$       &   $5.555$         &    $6.720$       &  $7.043$   \\
\hline 
 \end{tabular}
\caption{Same as Table \ref{diff_sph} but for the ground state configuration.}
\label{diff_gs} 
\end{table}

\begin{table}[!] \centering
 \begin{tabular}{|c|c|c|c|c|c|c|}
\hline	
     1$^{st}$ barrier    &  $^{216}$Th & $^{222}$Th        & $^{226}$Th        & $^{230}$Th        & $^{236}$U         & $^{240}$Pu  \\
\hline
 $\Delta {\rm E}_{{\rm HFB}}$       &    $7.272$  &  $7.474$    &   $7.568$       &  $7.888$         &       $7.425$    &  $7.141$  \\
 $\Delta {\rm E}_{{\rm MF}}$   & $1.456$     &  $2.221$    &   $-1.422$      &  $5.365$         &    $4.596$       &   $3.440$   \\
 $\Delta {\rm E}_{{\rm pair}}$ &  $0.293$    &  $-0.577$   &   $2.341$       &  $-3.574$         &    $-2.903$     &  $-1.824$    \\
 ${\rm E}_{{\rm TS}}$        &    $5.522$  &  $5.830$    &   $6.649$       &  $6.098$         &    $5.732$       & $5.525$ \\
\hline  
 \end{tabular}
\caption{Same as Table \ref{diff_sph} but for the first barrier configuration.}
\label{diff_bar1} 
\end{table}

\begin{table}[!] \centering
 \begin{tabular}{|c|c|c|c|c|c|c|}
\hline	
     2$^{nd}$ well          &  $^{216}$Th & $^{222}$Th        & $^{226}$Th        & $^{230}$Th        & $^{236}$U         & $^{240}$Pu  \\
\hline
 $\Delta {\rm E}_{{\rm HFB}}$        &   $7.537$   &    $7.638$        &     $7.982$      &   $7.572$         &      $7.977$     &  $7.852$  \\
 $\Delta {\rm E}_{{\rm MF}}$   &  $1.976$    &    $4.455$        &   $5.448$        &    $1.884$        &    $1.645$       &   $-2.561$   \\
 $\Delta {\rm E}_{{\rm pair}}$ &   $-0.335$  &    $-2.290$       &    $-3.387$      &    $-0.120$       &    $0.480$       &   $4.700$   \\
 ${\rm E}_{{\rm TS}}$        &  $5.895$    &    $5.472$        &    $5.920$       &     $5.807$       &    $5.852$       &  $5.713$   \\
\hline  
 \end{tabular}
\caption{Same as Table \ref{diff_sph} but for the second well configuration.}
\label{diff_swell} 
\end{table}

For the spherical configuration, one observes that $\Delta {\rm 
E}_{{\rm HFB}}$ is systematically positive. The same conclusion emerges 
when inspecting the ground state configurations as well as the first 
hump and the second well configurations. These nuclei are predicted 
less bound with the D1ST2a interaction by several MeV, which indicates 
that the proton-neutron part of the tensor term is the source of the 
global effect.
The analysis of the variation of $\Delta {\rm E}_{{\rm MF}}$ at the 
spherical point shows that it is systematically negative, which points 
out to a Hartree-Fock type mean-field which is more bound with the 
D1ST2a interaction by several MeV. This can be explained by the 
shifting of a few single-particle orbitals in presence of the tensor 
term around the Fermi level. This shifting produces a variation of the 
pairing energies $\Delta {\rm E}_{{\rm pair}}$ which is, in turn, 
systematically positive and larger in absolute value. As the pairing 
strength is identical for both interactions, one deduces that the 
rearrangement of the single-particle spectrum in presence of the tensor 
term tends to reduce the pairing contribution for spherical 
configurations in these nuclei. Finally, ${\rm E}_{{\rm TS}}$ is found 
positive. One notes the almost zero value obtained for $^{236}$U which 
is spin saturated in protons.

At the ground state deformation, which are prolate for five nuclei 
($^{216}$Th is excluded as its ground state is spherical), the detailed 
analysis of $\Delta {\rm E}_{{\rm HFB}}$ leads to opposite observations 
for $\Delta {\rm E}_{{\rm MF}}$ and $\Delta {\rm E}_{{\rm pair}}$. The 
mean-field is less bound with the D1ST2a interaction but the pairing 
energy is stronger. The contribution of the tensor term  ${\rm E}_{{\rm 
TS}}$ is always positive and larger than the one obtained at the 
spherical point.

At the deformations of the first hump, the variation $\Delta {\rm 
E}_{{\rm HFB}}$ is essentially dominated by the contribution ${\rm 
E}_{{\rm TS}}$. Both the quantities $\Delta {\rm E}_{{\rm MF}}$ and  
$\Delta {\rm E}_{{\rm pair}}$ have strongly decreased, in absolute 
value, in comparison with the two previous cases. No general trend is 
obtained for their signs as they depend on the nucleus.

At the deformations of the second well, the contribution ${\rm E}_{{\rm 
TS}}$ takes rather similar value for the six nuclei and it is still 
large. For the Thorium isotopes, $\Delta {\rm E}_{{\rm MF}}$ and 
$\Delta {\rm E}_{{\rm pair}}$ keep the same sign, associated with a 
less bound mean-field and stronger pairing correlations with the D1ST2a 
interaction. For $^{236}$U, $\Delta {\rm E}_{{\rm MF}}$, $\Delta {\rm 
E}_{{\rm pair}}$ and ${\rm E}_{{\rm TS}}$ are found positive for the 
D1ST2a interaction, which can be interpreted as a global repulsion. For 
$^{240}$Pu, the tensor term induces a more bound mean-field and a 
decrease of the pairing correlations. 

One concludes that, even though $\Delta {\rm E}_{{\rm HFB}}$ and ${\rm 
E}_{{\rm TS}}$ are found systematically positive for the six nuclei in 
the four states considered (spherical, ground state, first hump and 
second well),
no general law emerges concerning the mean-field and the pairing 
contributions except for the fact that they have in general  opposite 
sign. The results are subtle and depend strongly on the shell structure 
around the Fermi levels.


\subsubsection{Symmetric and asymmetric fission paths in $^{230}$Th, 
$^{226}$Th, $^{222}$Th and $^{216}$Th isotopes}

After these global comments on the axial deformation properties with 
and without parity breaking in even-even $^{216-232}$Th isotopes, one 
details now the symmetric and the asymmetric paths up to scission with 
both full 2D \{Q$_{20}$, Q$_{30}$\} potential energy surfaces (PES) and 
the associated 1D potential energy curves (PEC). Calculations have been 
done for the $^{230}$Th, $^{226}$Th and $^{222}$Th isotopes which sign 
experimentally the asymmetric to symmetric fission transition, as 
explained previously. We have kept also the $^{216}$Th isotope because 
of its N=126 magic neutron number even though not experimentally 
measured. Moreover, all along the paper, we have kept the pre-scission configurations in 
the considered collective variable space as far as possible in deformation and left aside the 
post-scission configurations which are characterized by the absence of matter between the two 
fragments, as it is usually done in fission studies (see for example Ref.\cite{Regnier1,Regnier2}). 
In the following, the post-scission configurations will be grouped under the term "fusion valley".

\begin{figure}[!] \centering
\begin{tabular}{c}
   \includegraphics[width=8.5cm]{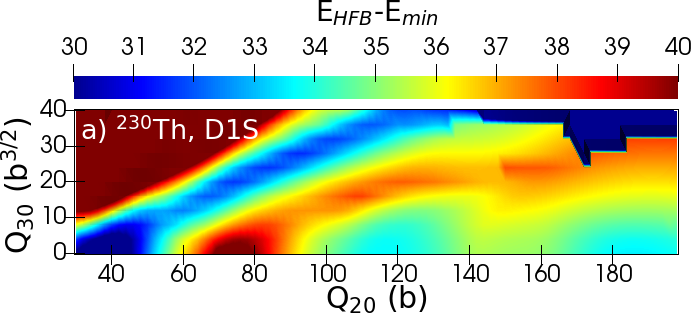} \vspace{0.3cm}\\ 
   \includegraphics[width=8.5cm]{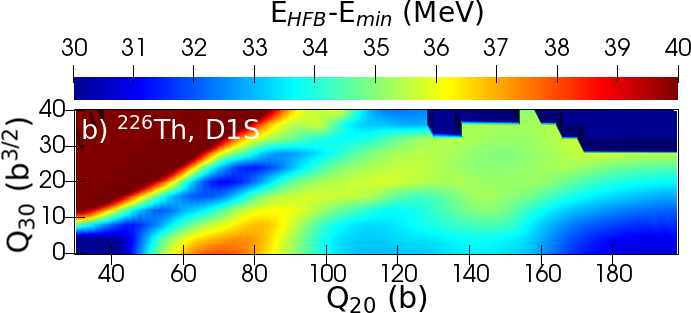} \vspace{0.3cm}\\ 
   \includegraphics[width=8.5cm]{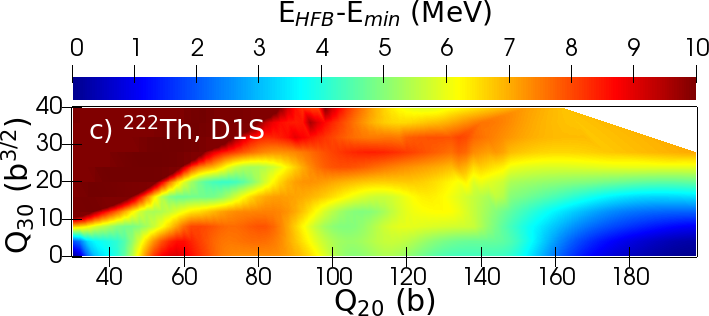} \vspace{0.3cm}\\ 
   \includegraphics[width=8.5cm]{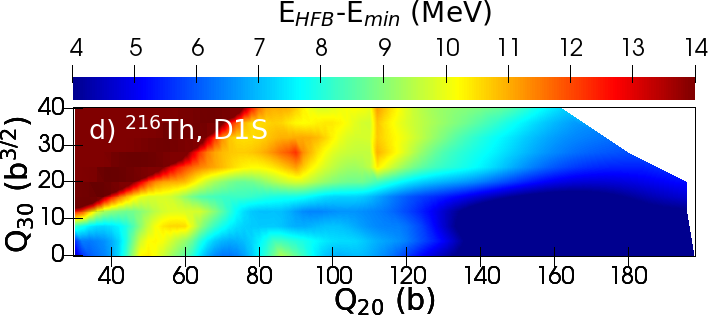} 
\end{tabular}
\caption{PES's for (a) $^{230}$Th (b) $^{226}$Th (c) $^{222}$Th and (d)$^{216}$Th as a function of the elongation Q$_{20}$ in b and the mass
asymmetry Q$_{30}$ in b$^{3/2}$. The color code indicates the HFB total energy normalized to the lowest value of 
the PES and spans a range of $10$~MeV. Calculations have been done with the D1S Gogny interaction. } 
\label{Ap3bb}
\end{figure}

\begin{figure}[!] \centering
\begin{tabular}{c}
   \includegraphics[width=8.5cm]{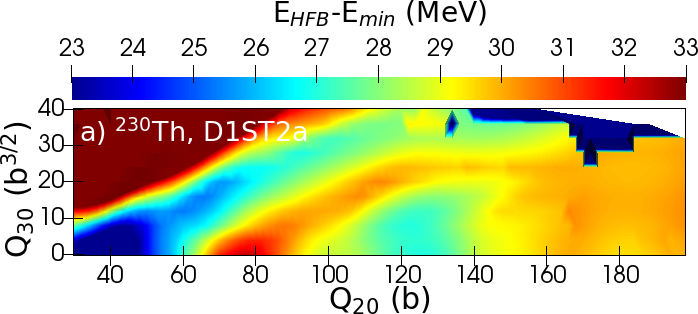} \vspace{0.3cm}\\ 
   \includegraphics[width=8.5cm]{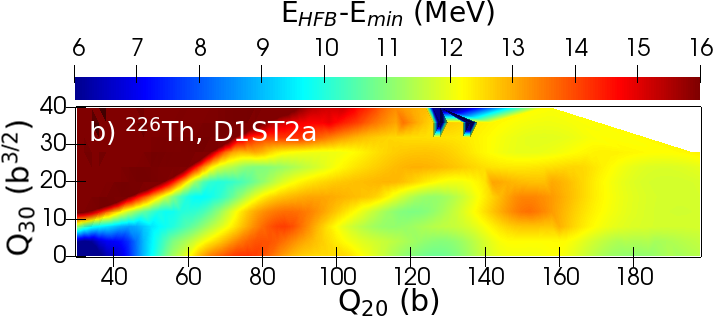} \vspace{0.3cm}\\ 
   \includegraphics[width=8.5cm]{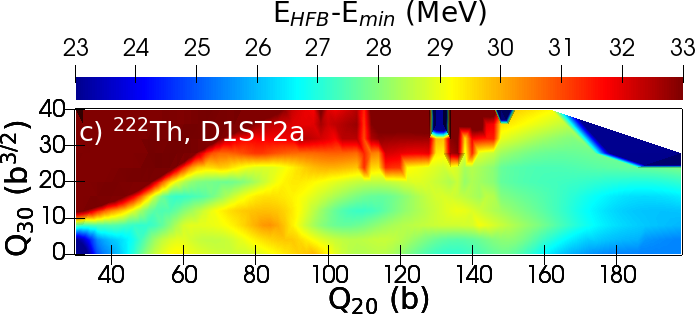} \vspace{0.3cm}\\ 
   \includegraphics[width=8.5cm]{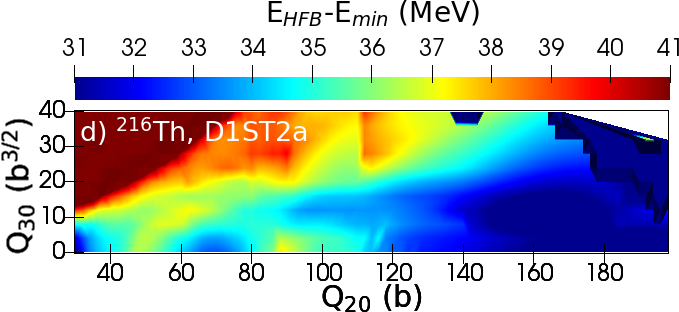} 
\end{tabular}
\caption{Same as Fig. \ref{Ap3bb} but for the D1ST2a Gogny interaction.} 
\label{Ap3bbc}
\end{figure}

\begin{figure} \centering
\begin{tabular}{c}  
   \includegraphics[width=8.5cm]{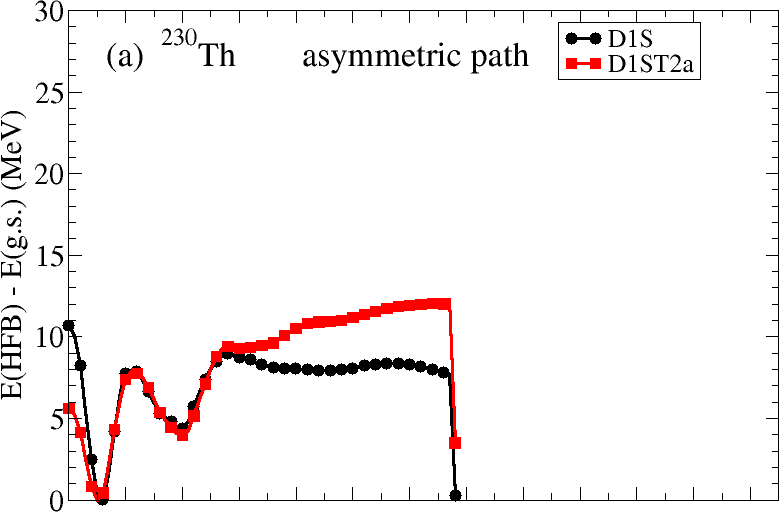}  \\ 
   \includegraphics[width=8.5cm]{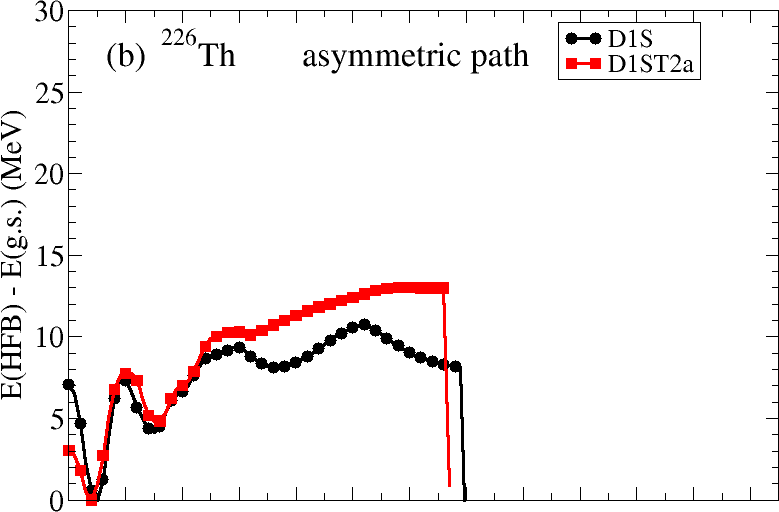}   \\ 
   \includegraphics[width=8.5cm]{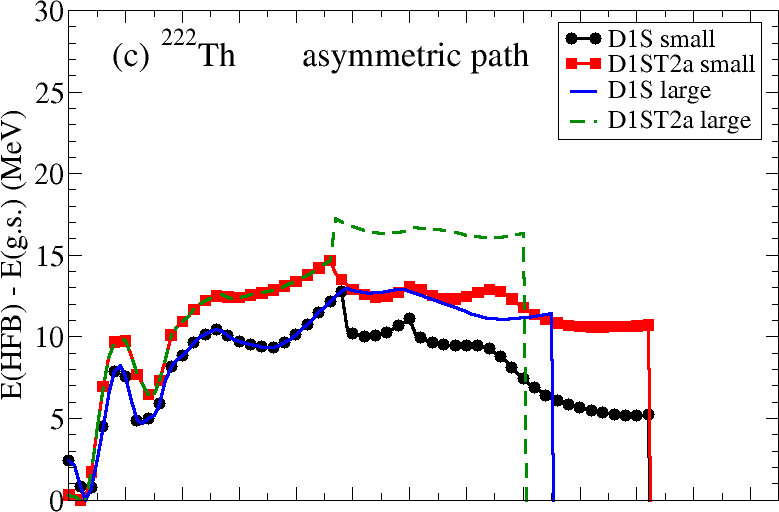} \\ 
   \includegraphics[width=8.5cm]{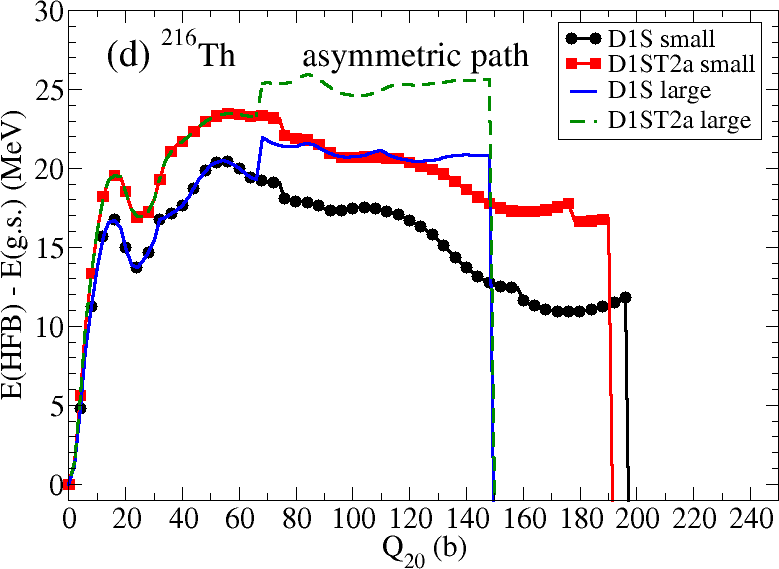}  
\end{tabular}
\caption{Asymmetric fission paths in (a) $^{230}$Th (b) $^{226}$Th (c) 
$^{222}$Th (d) $^{216}$Th isotopes calculated with the HFB 
approximation. Results correspond to D1S (full black circles) and 
D1ST2a (full red squares) Gogny interactions. The blue and the green 
curves are the large asymmetry path for the D1S and the D1ST2a 
interactions. See text for explanations. Energies are expressed in 
MeV.} 
\label{Ap3c}
\end{figure}

\begin{figure} \centering
\begin{tabular}{c}  
\includegraphics[width=8.5cm]{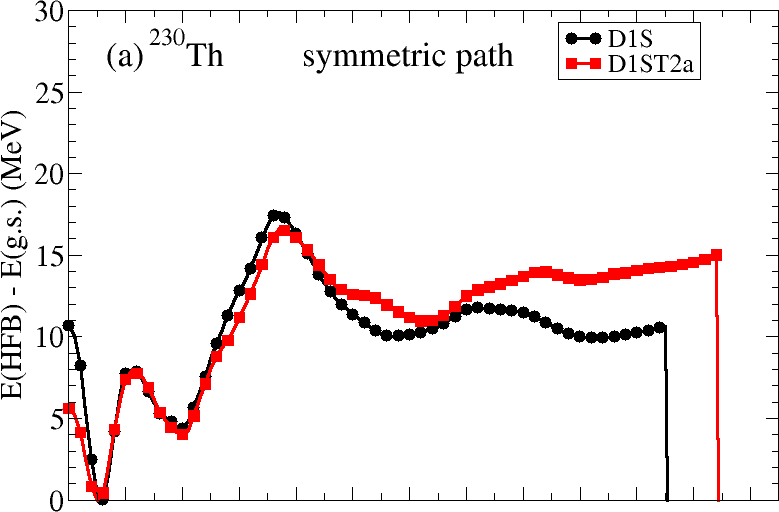} \\ 
\includegraphics[width=8.5cm]{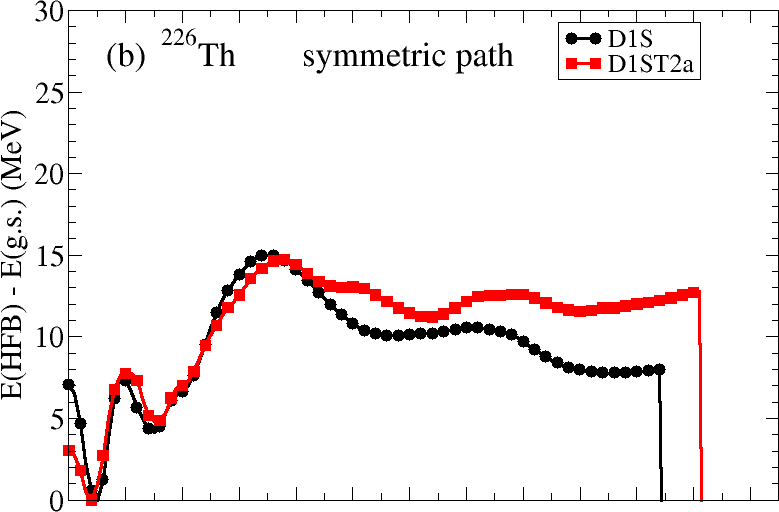} \\ 
\includegraphics[width=8.5cm]{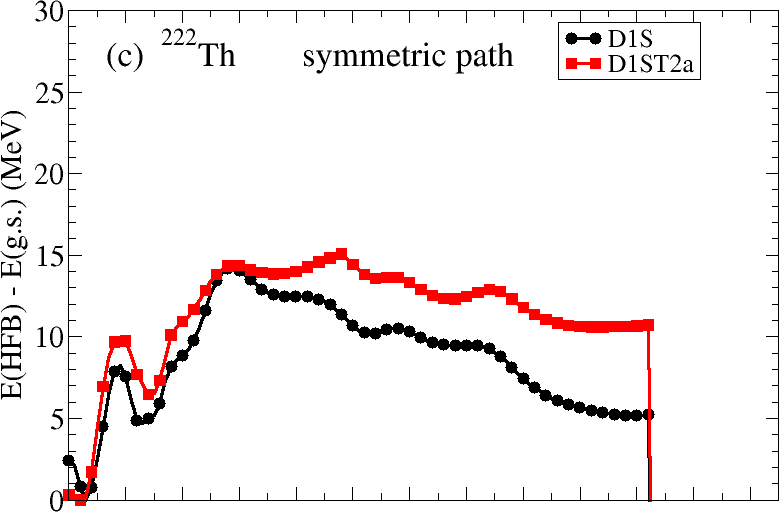} \\ 
\includegraphics[width=8.5cm]{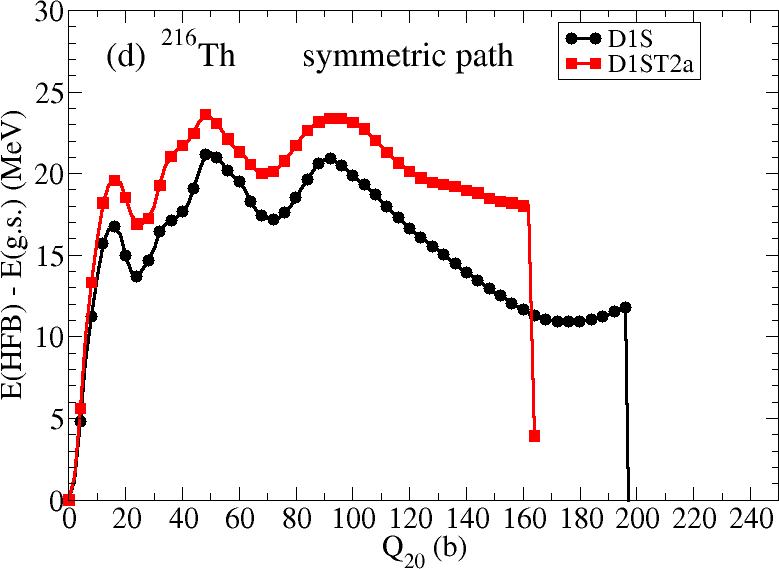} 
\end{tabular}
\caption{Symmetric fission paths in (a) $^{230}$Th (b) $^{226}$Th (c) 
$^{222}$Th (d) $^{216}$Th  isotopes calculated with the HFB 
approximation. Results correspond to the D1S (circles) and D1ST2a 
(squares)  Gogny interactions. Energies are expressed in MeV.} 
\label{Ap3d}
\end{figure}

\begin{figure} \centering
\begin{tabular}{c}
   \includegraphics[width=8.5cm]{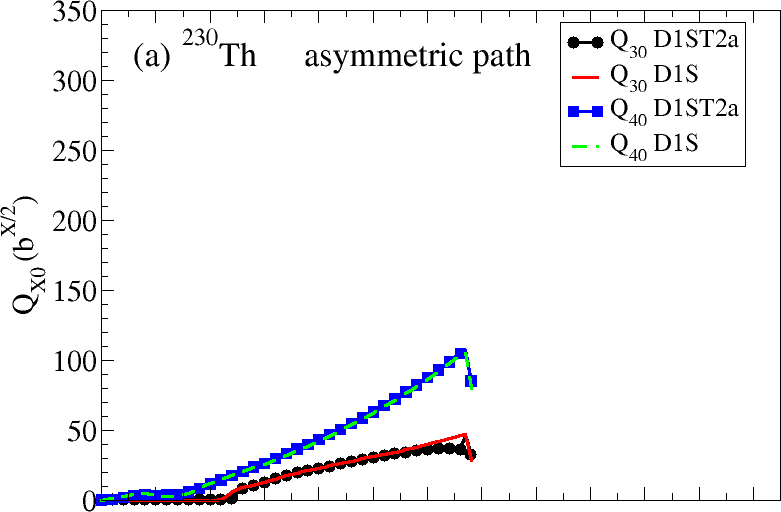} \\   
   \includegraphics[width=8.5cm]{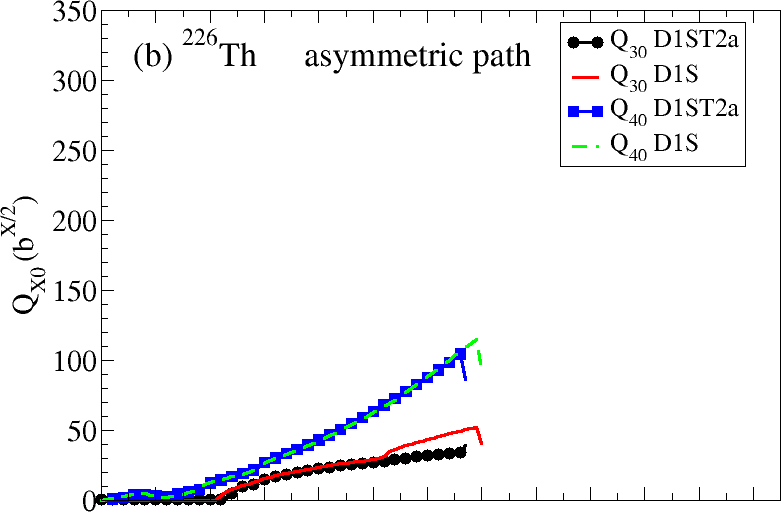}  \\ 
   \includegraphics[width=8.5cm]{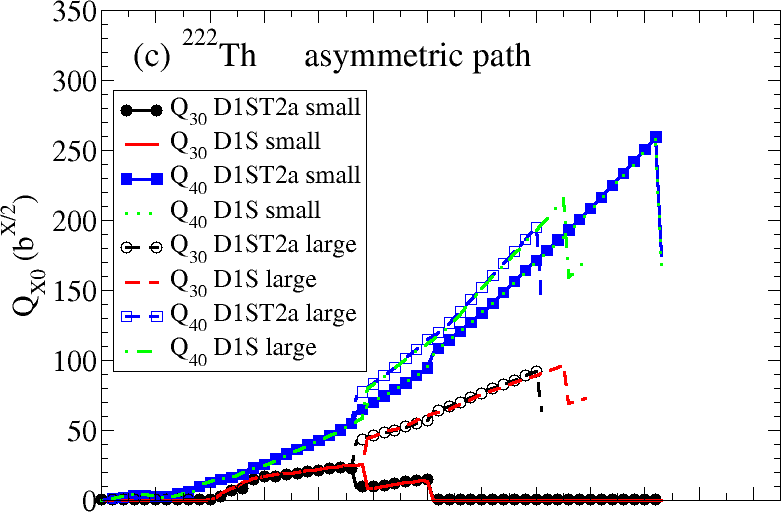} \\  
   \includegraphics[width=8.5cm]{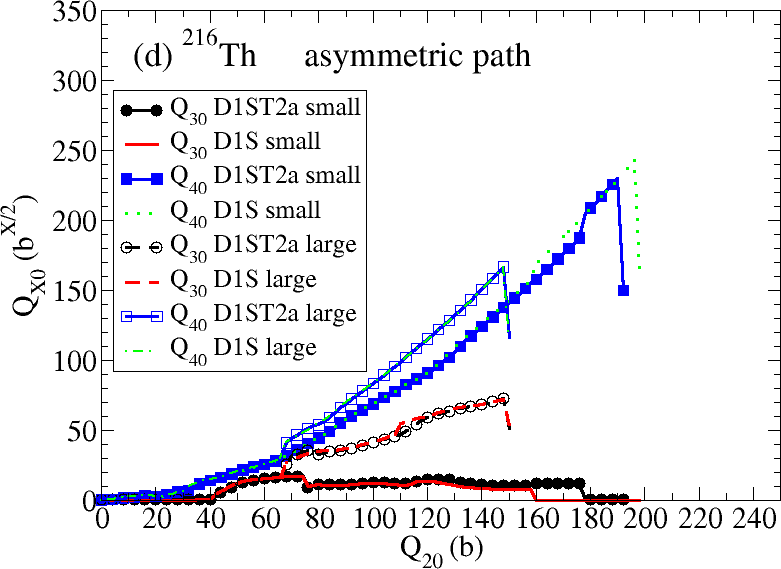} 
\end{tabular}
\caption{Evolution of Q$_{30}$ and Q$_{40}$ collective variables along 
the asymmetric path in (a) $^{230}$Th (b) $^{226}$Th (c) $^{222}$Th 
(d) $^{216}$Th isotopes calculated with the HFB 
approximation. Results are provided for both the D1ST2a and the D1S  
Gogny interactions. When they exist, the results for the small and the 
large asymmetry paths are shown.} 
\label{Ap4c}
\end{figure}

\begin{figure} \centering
\begin{tabular}{c}
\includegraphics[width=8.5cm]{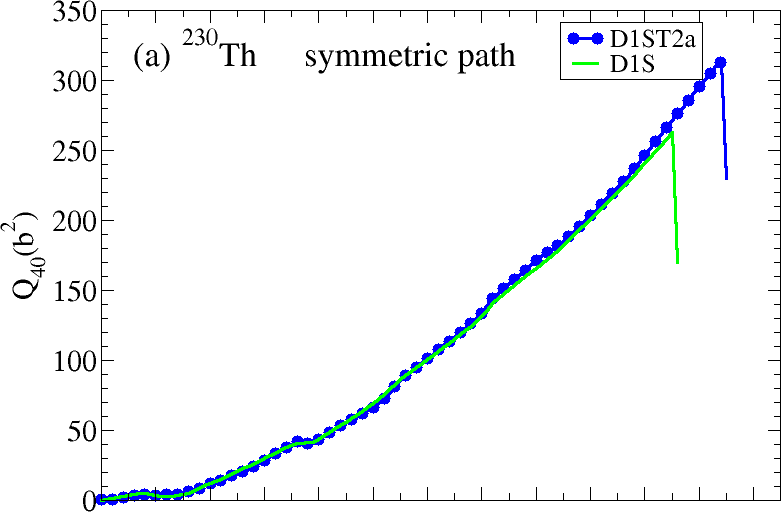} \\   
\includegraphics[width=8.5cm]{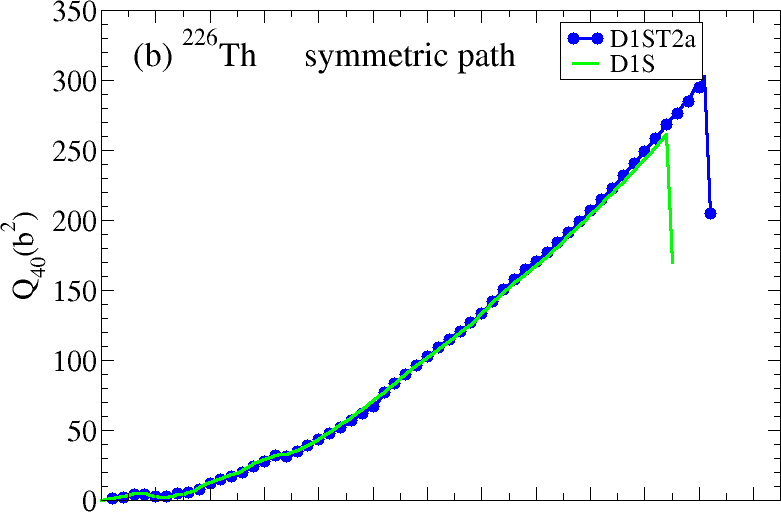} \\  
\includegraphics[width=8.5cm]{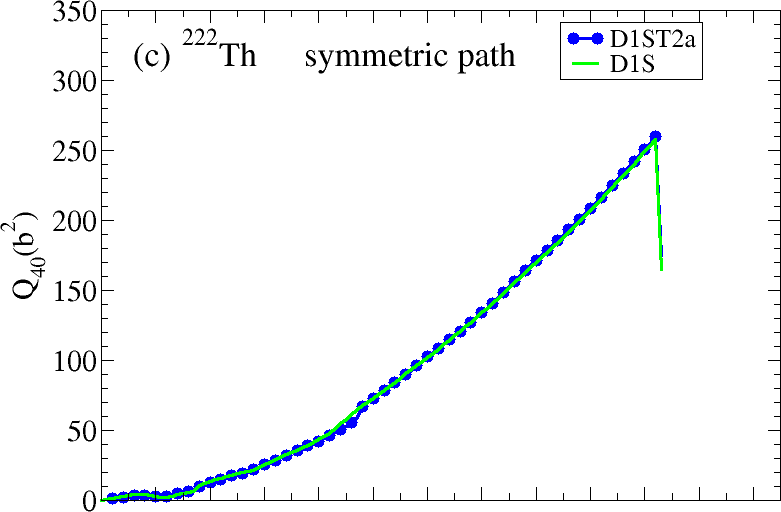} \\  
\includegraphics[width=8.5cm]{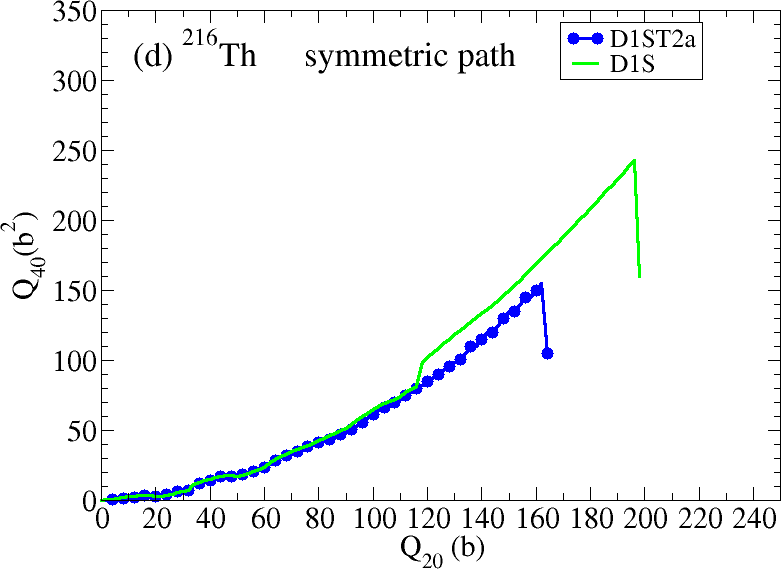} 
\end{tabular}
\caption{Evolution of Q$_{40}$ collective variable symmetric path in 
(a) $^{230}$Th (b) $^{226}$Th (c) $^{222}$Th (d)
$^{216}$Th isotopes calculated with the HFB approximation. Results are 
provided for both the D1ST2a and the D1S Gogny interactions.} 
\label{Ap4d}
\end{figure}

On Fig. \ref{Ap3bb}, the PES, obtained using the D1S Gogny interaction, 
for the four isotopes considered are plotted. The x-axis corresponds to 
the elongation Q$_{20}$ which ranges between $30$~b (around the 
deformation of the second well) and $200$~b. The y-axis represents the 
asymmetry Q$_{30}$ which varies between $0$~b$^{3/2}$ and 
$40$~b$^{3/2}$. The color code ranges over $10$~MeV for all the panels 
(a), (b) , (c) and (d). It represents the energy difference between the 
HFB total binding energy for given values of Q$_{20}$ and Q$_{30}$, and 
the lowest HFB value obtained in the existing PES. For convenience a 
Delaunay triangulation has been performed as in \cite{RegCPC18} for all 
the 2D PES. The results with the D1ST2a interaction are displayed in 
Fig. \ref{Ap3bbc}. 
The corresponding 1D asymmetric and symmetric paths 
are drawn in Fig. \ref{Ap3c} and Fig. \ref{Ap3d}, respectively, 
according to the collective variable Q$_{20}$ between $0$~b and 
$250$~b. The total HFB energy E$_{HFB}$ has been renormalized to the 
ground state total energy E$_{g.s.}$. Results are indicated for both 
the D1S (black full circles) and the D1ST2a (red full squares) 
interactions. The evolution of the associated collective variables 
Q$_{30}$ (for the parity breaking paths) and Q$_{40}$ are shown on Fig. 
\ref{Ap4c} and Fig. \ref{Ap4d}.

For the $^{230}$Th isotope, Fig. \ref{Ap3bb} (a), one observes the 
existence of an asymmetric path which starts around Q$_{20}\simeq 50$~b 
(see Fig. \ref{Ap4c} (a)) and leads to static HFB configurations with a 
large asymmetry. This path is clearly the lowest in energy. It seems to 
be rather flat (with a slight decrease of the energy for increasing 
Q$_{20}$) and displays scissionned configurations around Q$_{20} \simeq 
\rm 137 ~b$ for Q$_{30}\simeq 40$~b$^{3/2}$ and Q$_{40} \simeq 
110$~b$^{2}$, as seen from Fig. \ref{Ap3c} (a). 
From Fig.\ref{Ap3bbc}(a), FIG.\ref{Ap3c} (a) and Fig. \ref{Ap4c}(a), one 
concludes that these observations hold also for the D1ST2a interaction, 
except that the energy of this large asymmetry path increases slowly 
from the second hump up to the scissionned configuration.

For comparison, as seen from Fig. \ref{Ap3bb} (a), Fig.\ref{Ap3d} (a) 
and Fig.\ref{Ap4d} (a), one obtains that the symmetric path is less 
favorable energetically because of the height of the second hump which 
is predicted to be $\simeq 18$~MeV ($\simeq 17$~MeV) for the D1S 
(D1ST2a) Gogny interaction. Moreover, one sees that the scissionned 
configuration, which defines in our case the exit point, is encountered 
at a much larger value of Q$_{20}$, around $\simeq211$~b for D1S and 
$\simeq 229$~b for D1ST2a, with a larger hexadecapole moment equal to 
Q$_{40} \simeq 250$~b$^{2}$ and $300$~b$^{2}$, respectively.

For the $^{226}$Th isotope, the path which leads to large asymmetry 
scission still exists with the D1S interaction, as seen in Fig. 
\ref{Ap3bb} (b). It starts around Q$_{20}=45$~b as indicated in Fig. 
\ref{Ap4c} (b). 
It presents a more pronounced third hump than in the 
$^{230}$Th isotope, around Q$_{20} \simeq 105$~b and Q$_{30} 
\simeq30$~b$^{3/2}$ which is easily identifiable in Fig. \ref{Ap3c} 
(b). The tensor term tends to increase by $\simeq 1$~MeV the height of 
the second hump and the rest of the large asymmetry path displays a 
continuous increase of the total energy up to the scissionned point, as 
already discussed for the $^{230}$Th isotope. The first scissionned 
configurations are obtained at Q$_{20} \simeq 140$~b with the D1S 
interaction and at a little smaller value for the D1ST2a interaction, 
namely 132b. Moreover, from Fig. \ref{Ap4c} (b), one sees that the 
values of Q$_{30}$ and Q$_{40}$ are very similar to the ones obtained 
at the exit point in the $^{230}$Th isotope.

In any case, comparing Fig. \ref{Ap3c} (b) with Fig. \ref{Ap3d} (b), 
one observes that for both interactions, the asymmetric path is again 
lower in energy. Indeed, even though the height of the second hump for 
the symmetric path is $\sim 15$~MeV for both interactions, which is 
lower than the one in $^{230}$Th isotope, it is still higher than the 
asymmetric one which is equal to $\sim 9$~MeV ($10$~MeV) for the D1S 
(D1ST2a) interaction. One adds that the symmetric exit points are 
characterized by very similar values than the ones obtained for the 
$^{230}$Th isotope but with a little decrease of Q$_{20}$.

For the $^{222}$Th isotope, the situation starts to be different. From 
Fig. \ref{Ap3bb} (c), one observes that this big asymmetry path gets 
clogged, which was already the case in the $^{230}$Th and $^{226}$Th 
isotopes with the D1ST2a interaction. At some point, around Q$_{20} 
\simeq 90$~b, it is no longer energetically favorable. This 
corresponds to the energy discontinuity observed in the red full squares 
and black full circles 
curves, Fig. \ref{Ap3c} (c). Here, the blue full line and the green dashed line 
correspond to the continuation of paths which lead to large asymmetry 
scission and which are clearly higher in energy by several MeV. At this 
deformation, it is  higher by $\sim 1.5$~MeV ($3$~MeV) with D1S 
(D1ST2a) interaction. The minimum energy principle implies a path with 
a smaller asymmetry around $13$~b$^{3/2}$, which is located in a new 
local minimum as observed in Fig. \ref{Ap3bb} (c). Elongating more and 
more the nucleus, one finds that the symmetric configuration is 
energetically favorable around Q$_{20} \simeq 120$~b, which is 
signaled by the energy discontinuity observed in Fig. \ref{Ap3c} (c) at 
this quadrupole deformation. The system continues along the symmetric 
path up to scission. The exit point appears at Q$_{20} \simeq 205$~b 
for both interactions, which is characteristic of the well-known super 
long symmetric fission mode, with Q$_{40} \simeq 260$~b. At this level, 
no sign of possible compact fission can be highlighted. The most 
important phenomenon obtained in $^{222}$Th isotope is the rebalancing 
of the heights of the symmetric and asymmetric second humps due to the 
tensor term, as seen from Fig. \ref{Ap3c} (c) and Fig. \ref{Ap3d} (c). 
This rebalancing is characterized by an increase of the height of the 
asymmetric second hump in presence of the tensor term, whereas the 
symmetric one is essentially unchanged. From a 1D energetic viewpoint, 
this renders the full symmetric path more probable (or less 
improbable!) when the tensor term is added, and therefore the symmetric 
fission mode.

For the $^{216}$Th isotope, the same type of mixed asymmetric-symmetric 
path manifests as the one found in the $^{222}$Th isotope (see Fig. 
\ref{Ap3bb} (d), Fig. \ref{Ap3bbc} (d), Fig. \ref{Ap3c} (d), Fig. 
\ref{Ap3d} (d), Fig. \ref{Ap4c} (d) and Fig. \ref{Ap4d} (d)). One 
obtains first a path which leads to large asymmetry scission from 
Q$_{20}\simeq 40$~b up to Q$_{20}\simeq 75$~b, then a path with small 
asymmetry characterized by Q$_{30} \simeq 10$~b$^{3/2}$ up to 
Q$_{20}\simeq 160$~b (178b) with the D1S (D1ST2a) interaction, followed 
by a symmetric path. In that case, the first symmetric scissionned point 
is obtained at Q$_{20} \simeq 198$~b for the D1S interaction, and a 
smaller value of 191b for the D1ST2a one. Now, looking at the pure 
symmetric path (Fig. \ref{Ap3d} (d)), one observes that, in the case of 
the D1ST2a interaction, the exit point is obtained at Q$_{20} \simeq 
162$~b which is different and much smaller than the value deduced from 
the mixed asymmetry path. This is an unusual short value for a 
symmetric scission, more characteristic of an asymmetric scission. As 
we will see later, this is a first theoretical hint of the existence of 
the compact fission mode that could be correlated with the observations 
of the SOFIA experiment \cite{SOFIA1,SOFIA2}. From Fig. \ref{Ap4d} (d), 
one sees that this mode is characterized also by a much smaller value 
of the hexadecapole moment Q$_{40}$ at the exit point, which is equal 
to $\sim 150$~b$^{2}$.

Finally, one observes a global and significant increase of the height 
of the barriers in the $^{216}$Th isotope which has a neutron magic 
number equal to 126, for both the symmetric and the asymmetric paths, 
whatever the interaction. One notes that the symmetric second hump 
height is lower than the asymmetric one in the case of D1ST2a. In 
conclusion, one sees that the tensor term plays a non-negligible role 
on the barrier height. Its behavior is a detailed one which acts 
differently on the symmetric and the asymmetric path. At this stage, 
within these 1D and 2D analysis made in terms of Q$_{20}$, Q$_{30}$ 
collective variables, no clear explanation is available concerning its 
role in the existence of the symmetric compact fission mode along the 
isotopic chain. It appears only in the $^{216}$Th isotope.

\subsection{Tensor term effect and symmetric compact scission - Role of the Q$_{40}$ collective variable}\label{resultsB}

\begin{figure} \centering
\begin{tabular}{c}
\includegraphics[width=6.0cm]{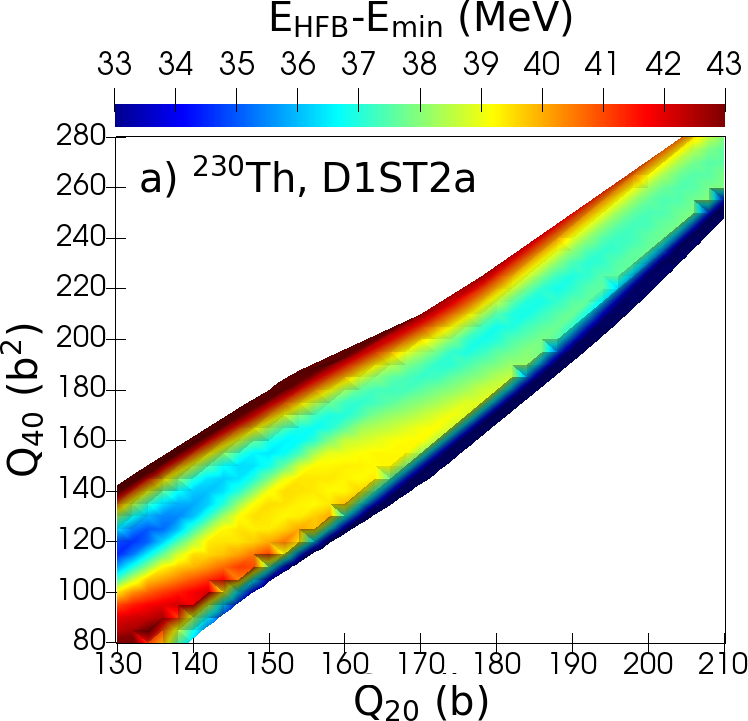} \vspace{0.2cm} \\ 
\includegraphics[width=6.0cm]{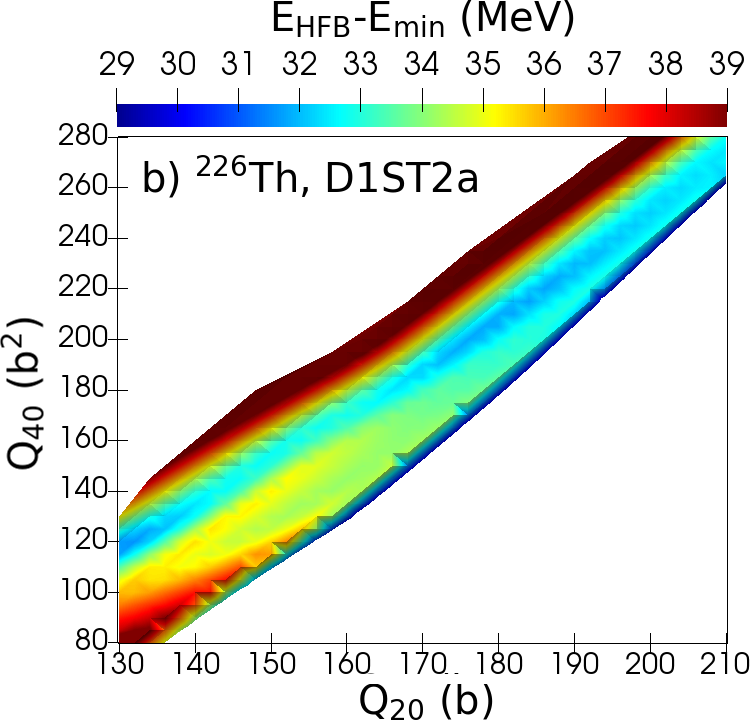} \vspace{0.2cm} \\ 
\includegraphics[width=6.0cm]{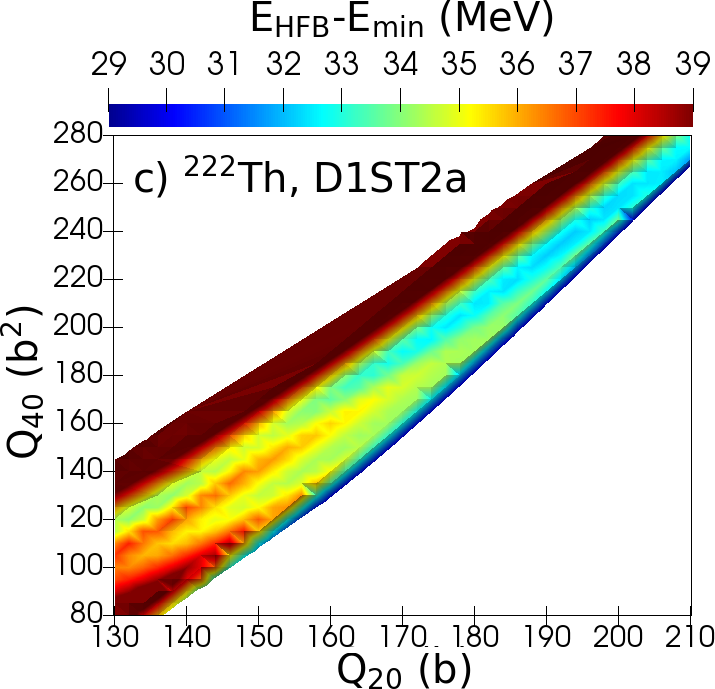} \vspace{0.2cm} \\ 
\includegraphics[width=6.0cm]{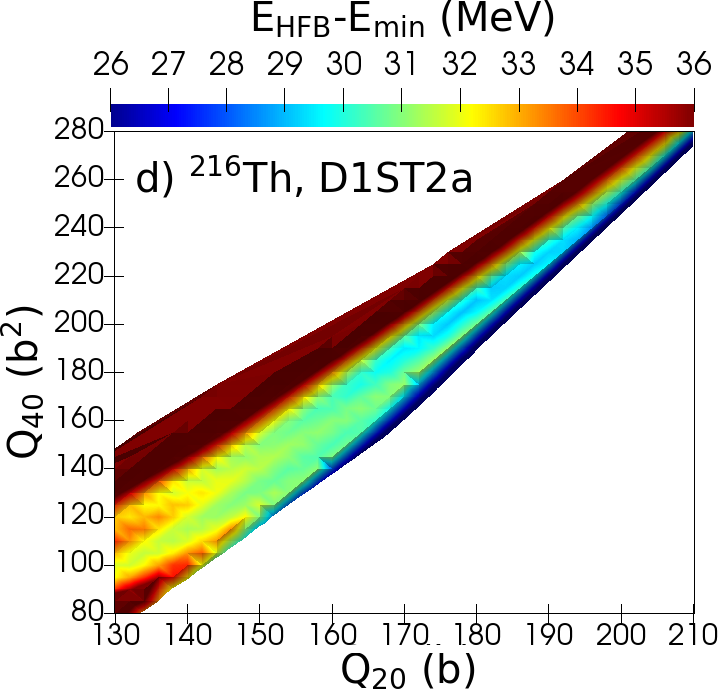} 
\end{tabular}
\caption{\{Q$_{20}$, Q$_{40}$\} potential energy surfaces associated with the symmetric fission path (Q$_{30}$=0) for (a) $^{230}$Th
(b) $^{226}$Th (c) $^{222}$Th and (d) $^{216}$Th. Calculations have been done with the D1ST2a interaction. Energies are expressed in MeV.} 
\label{collQ404}
\end{figure}

\begin{figure} \centering
\begin{tabular}{c}
\includegraphics[width=6.0cm]{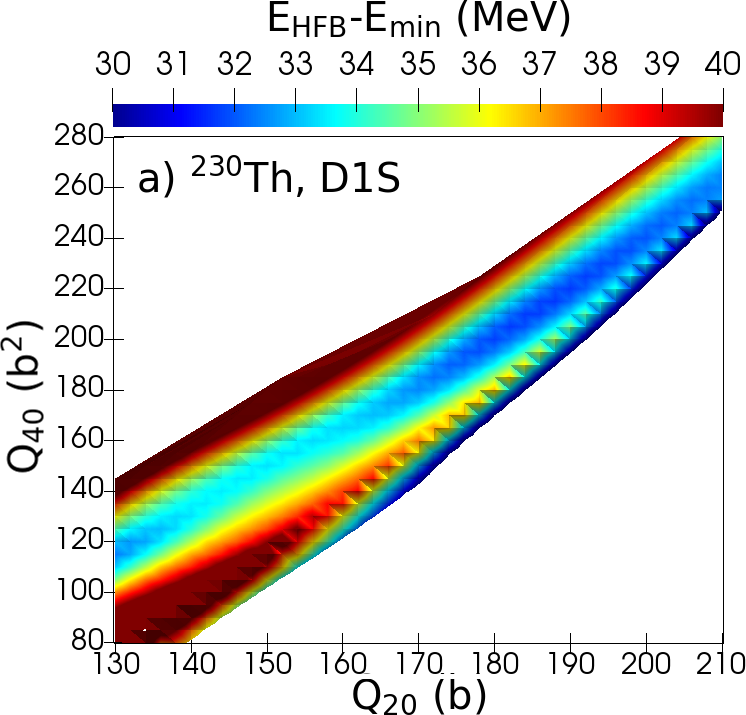}  \vspace{0.2cm} \\ 
\includegraphics[width=6.0cm]{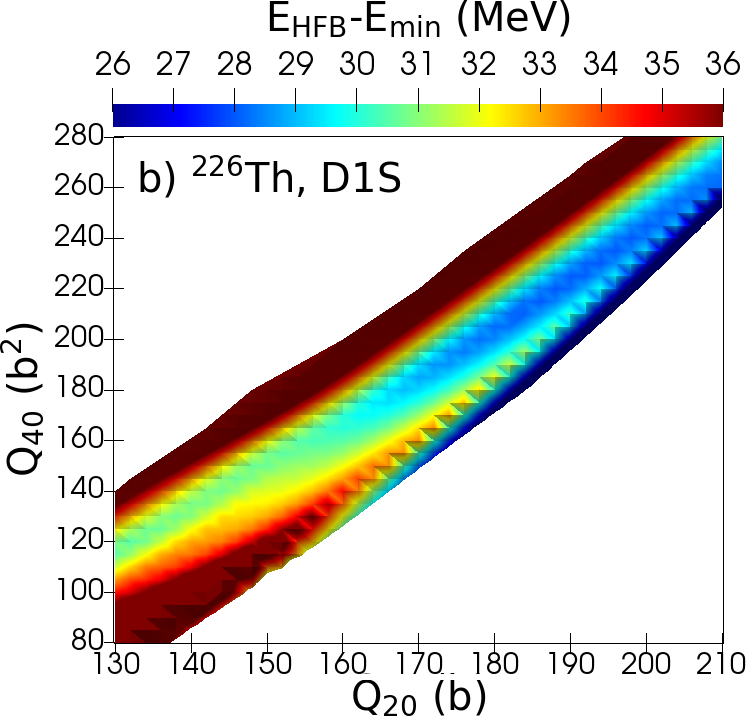}  \vspace{0.2cm} \\ 
\includegraphics[width=6.0cm]{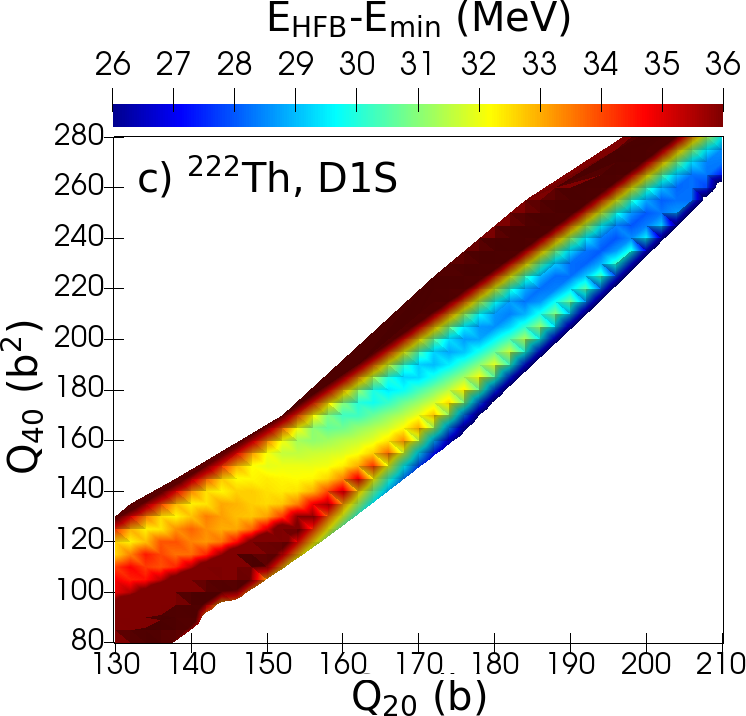}  \vspace{0.2cm} \\ 
\includegraphics[width=6.0cm]{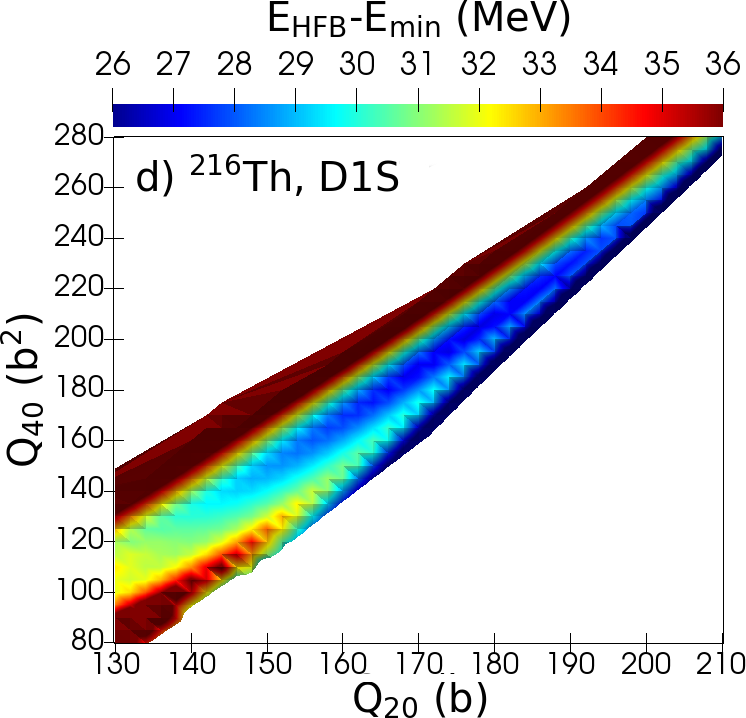} 
\end{tabular}
\caption{Same as Fig. \ref{collQ404} but for the D1S interaction.} 
\label{collQ403}
\end{figure}

In this part, the potential role of the Q$_{40}$ collective variable to 
explain the existence of the symmetric compact fission mode in light 
Thorium isotopes is investigated. This possibility has been suggested 
by the results obtained for the symmetric path in the $^{216}$Th 
isotope with the D1ST2a interaction, for which the first scissionned 
configuration is characterized by both a much smaller value of Q$_{20}$ 
and Q$_{40}$ in comparison with the other isotopes. To perform this 
analysis, the 2D PES's using the \{Q$_{20}$ and Q$_{40}$\} collective 
variables, have been calculated with both interactions. 
The quadrupole 
moment Q$_{20}$ ranges from $130$~b up to $300$~b and the hexadecapole one 
Q$_{40}$ from $90$~b$^{3/2}$ up to $300$~b$^{3/2}$. 
The results are 
shown on Fig. \ref{collQ404} (Fig. \ref{collQ403}) for the D1ST2a 
interaction (D1S) for the four selected Thorium isotopes.

\begin{figure} \centering
\begin{tabular}{cc}
\includegraphics[width=8.5cm]{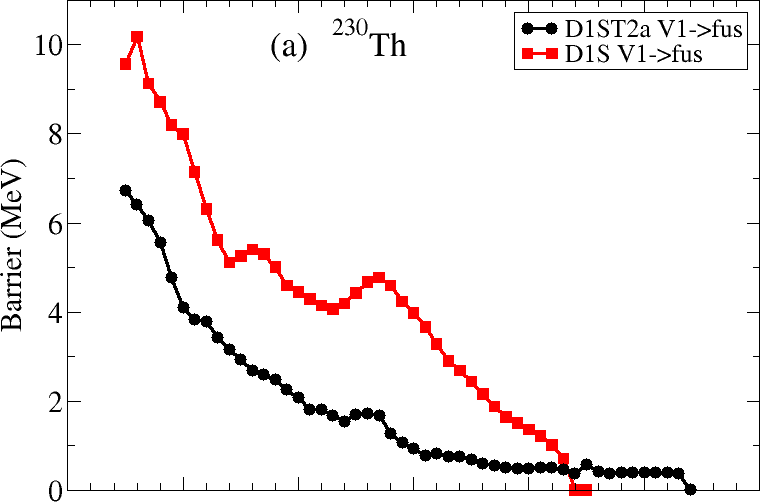} \\
\includegraphics[width=8.5cm]{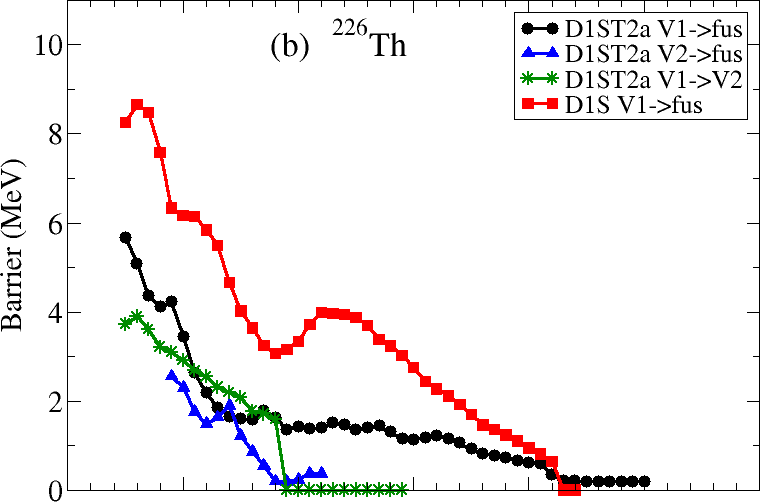} \\
\includegraphics[width=8.5cm]{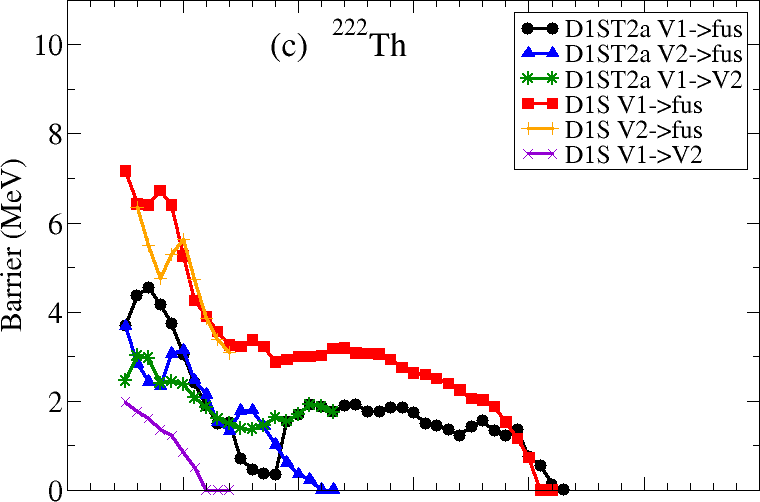} \\
\includegraphics[width=8.5cm]{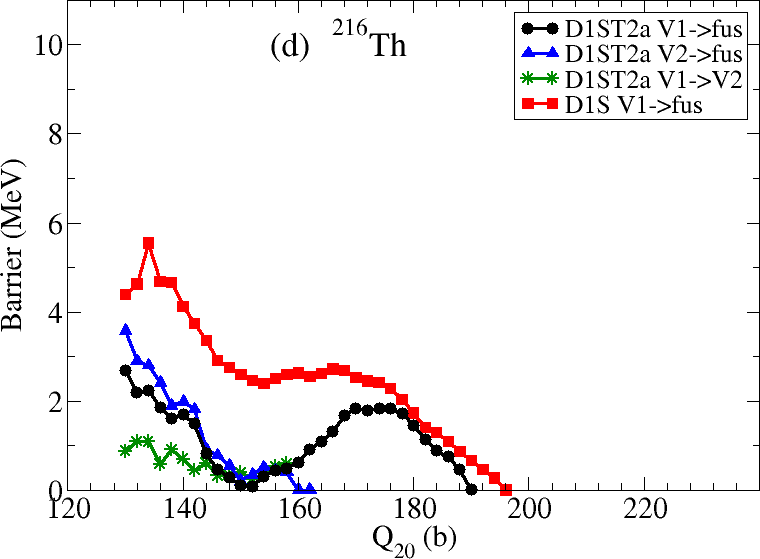}
\end{tabular}
\caption{Evolution of barrier heigths in (a) $^{230}$Th, (b) $^{226}$Th, (c) $^{222}$Th and (d) $^{216}$Th calculated 
with the D1S and D1ST2a Gogny interactions.
Energies are expressed in MeV.} 
\label{barrierQ40a}
\end{figure}

Concerning  the $^{230}$Th isotope, the calculation with D1ST2a shows a 
unique valley as can be seen in Fig. \ref{collQ404} (a). On the right
hand side of this main valley, called V1 in the following, one notes 
the existence of a kind of small plateau colored in yellow and located 
a few MeV above the bottom of the valley V1. For comparison, in the 
case of the D1S interaction (see Fig. \ref{collQ403} (a)), only a 
well-defined valley exists. The exit point is characterized by Q$_{20} 
\simeq 230$~b ($210$~b) and Q$_{40} \simeq 315$~b$^{2}$ ($255$~b$^{2}$) 
for the D1ST2a (D1S) interaction. In order to analyze in more details 
these results,  the evolution of the barrier heights between the 
fission V1 and the fusion (called "fus") valleys as a function of 
Q$_{20}$ is shown in Fig. \ref{barrierQ40a} (a) for both the D1ST2a 
(full black circles) and the D1S (full red squares) interactions. These 
barrier heights have been defined as the values deduced from 
transversal slices to the path which follows the bottom of the valley. 
At the beginning, around Q$_{20} \simeq 130$~b, the barrier height is 
around $7$~MeV for the D1ST2a interaction. Then, increasing the 
elongation Q$_{20}$ of the nucleus, it decreases and reaches a value 
which is lower than $1$~MeV around $180$~b. Finally, it remains stable up 
to $\simeq 225$~b and disappears around $230$~b at the exit point. With 
the standard D1S interaction, the value of the barrier is 
systematically higher by $2-3$~MeV along the symmetric path. Only at 
the end, its value decreases rapidly and goes to zero at a value of 
Q$_{20}$ slightly smaller, around $\simeq 210$~b. From these results, 
we conclude that the tensor term tends to decrease by several MeV the 
height of the V1 to fusion barrier.

For the $^{226}$Th isotope, the difference between the patterns 
obtained with the D1ST2a and the D1S interactions begins to intensify. 
The main valley V1 existing in the $^{230}$Th isotope is still there. 
However, as seen from Fig. \ref{collQ404} (b), the plateau changes into 
a kind of proto-valley, called V2 in the following. It appears around 
Q$_{20} \simeq 140$~b for a smaller value of Q$_{40}$ which 
characterizes the valley V1, around $\simeq 110$~b$^{2}$. The evolution 
of the values of the different transverse barrier heights is reported 
on Fig. \ref{barrierQ40a} (b). Concerning the principal valley V1, the 
barrier heights "V1 $\to$fus" (full black circles for D1ST2a and full 
red squares for D1S) are of the same order of magnitude as the ones 
obtained in $^{230}$Th, even a little smaller. Their relative behavior 
is similar with a cancellation of the barriers around Q$_{20} \simeq 
222$~b for D1ST2a and $\simeq 208$~b for D1S. Concerning the barrier 
between the principal valley V1 and the proto-valley V2 (green stars) 
which exists with the D1ST2a interaction and which is referred to as 
"V1$\to$V2", its height is equal to $\simeq 4$~MeV at its nascence 
around Q$_{20} \simeq 140$~b and decreases by $2$~MeV up to Q$_{20} 
\simeq 157$~b where V2 suddenly disappears. At this elongation, the 
barrier between the principal valley V1 and the fusion valley "fus" is 
still $\sim2$~MeV.

The value of the barrier between the proto-valley V2 and the fusion 
valley, named "V2$\to$fus" (full blue triangles), starts at $\simeq 
2.5$~MeV, decreases regularly and cancels around Q$_{20}\simeq 165$~b. 
Please note that its associated exit point is characterized by smaller 
values of Q$_{20}$ and Q$_{40}$ than the ones of the main valley V1, 
$\simeq 166$~b and $\simeq 150$~b$^{2}$, respectively. It signals the 
possible existence of a symmetric compact fission mode. The main 
question which remains to answer is the possibility of feeding the 
proto-valley which is located at a couple of MeV above V1 in this 
isotope. One can invoke two possibilities in the adiabatic hypothesis: 
either by tunnel effect or by excitation of a transverse mode which is 
the most probable mechanism. Another possibility would be to populate 
the valley through individual quasi-particle excitation with the 
available energy acquired after the saddle point.

In the case of the $^{222}$Th isotope, the observations made for 
$^{226}$Th are confirmed. The proto-valley transforms into a 
well-identified second valley which appears around Q$_{20} \simeq 
130$~b and Q$_{40} \simeq 100$~b$^{2}$, as seen in Fig. \ref{collQ404} 
(c). For comparison, the associated value of Q$_{40}$ for the V1 
valley is $\simeq 120$~b$^{2}$, which corresponds to an increase of 
20\%. These values associated with the new valley V2 are also smaller 
than the ones of the proto-valley in the $^{226}$Th isotope. We note 
that the new valley V2 is higher in energy than the valley V1, but it 
is lower in energy than the proto-valley V2 found in $^{226}$Th 
isotope. Its exit point is found at Q$_{20} \simeq 166$~b and Q$_{40} 
\simeq 150$~b$^{2}$. The exit point of the principal valley V1 is 
characterized by Q$_{20} \simeq 208$~b and Q$_{40} \simeq 265$~b$^{2}$, 
which are much larger values. Thus, the V1 and V2 valleys define two 
distinct modes in the symmetric path for $^{222}$Th: the classic super 
long mode from V1 and a new compact mode induced by the tensor term of 
the nuclear interaction.

For the D1S interaction,  a kind of "tilted plateau" appears on the 
right side as can be seen in Fig. \ref{collQ403} (c). It clearly ends in 
the principal valley V1. The structure of this "tilted plateau" seems 
to be different from the structure of a valley. In particular, it 
disappears in the $^{216}$Th isotope, as discussed below. The exit 
point corresponds to a well-elongated fission with Q$_{20} \simeq 
206$~b and Q$_{40} \simeq 255$~b$^{2}$.
The evolution of the barrier heights can be seen on Fig. 
\ref{barrierQ40a} (c). Because of the existence of the "tilted plateau" 
with the D1S interaction, we have drawn for both interactions the 
barriers "V1$\to$fus", "V2$\to$fus" and "V1$\to$V2", where V2 
represents the second valley in the case of D1ST2a and the tilted 
plateau for D1S. Those barriers are the energy difference between the 
bottom of the first valley and the crest separating the two valleys.

When considering the "V1$\to$fus" barrier obtained with the D1ST2a 
interaction (full black circles), one has to be careful with the 
interpretation and has also to consider the barrier "V1$\to$V2" (full 
blue triangles) which separates both valleys. Indeed, the principal 
valley is not connected directly to the fusion valley at the beginning 
of the path up to the exit point of the valley V2. Around Q$_{20} 
\simeq 130$~b, the barrier height "V1$\to$fus" is equal to $\simeq 
4$~MeV, which is lower by $1.5$~MeV in comparison with the one obtained in 
$^{226}$Th. Then, it quickly decreases  up to Q$_{20} \simeq 156$~b 
where it reaches a small value of $\simeq 400$~keV. However, as 
previously mentioned, the "V1$\to$V2" barrier height in this 
deformation region is around $1.7$~MeV. Moreover, around Q$_{20} \simeq 
158$~b, both barriers "V1$\to$V2" and "V1$\to$fus" become mixed up. The 
"V1$\to$V2" barrier disappears at Q$_{20} \simeq 168$~b (the exit point 
of the valley V2). Only the barrier "V1$\to$fus" exists for larger 
values of Q$_{20}$. Its height stay more or less constant up to Q$_{20} 
\simeq 200$~b and is equal to $\simeq 1.7$~MeV. Then, it decreases 
rapidly and goes down to zero at Q$_{20} \simeq 208$~b. Finally, one 
observes that the "V1$\to$V2" barrier height is not changing too much, 
being equal to $\simeq 3$~MeV for the smallest Q$_{20}$ values and 
$\simeq 1.7$~MeV for larger ones. Concerning the "V2$\to$fus" barrier 
height, after a fluctuation around $\simeq$3~MeV for the smallest values 
of Q$_{20}$, it decreases and disappears at Q$_{20} \simeq 165$~b, a 
value compatible with the compact fission mode.

For the D1S interaction, the situation is different because of the 
presence of the tilted plateau. First of all, the height of the tilted 
plateau to the bottom of the principal valley V1, called "V1$\to$V2", 
changes rapidly (purple crosses). Around Q$_{20} \simeq 130$~b, it is 
equal to $\simeq 2.0$~MeV. Around the elongation Q$_{20} \simeq 145$~b, 
it disappears. For larger deformations, the principal valley V1 is 
directly connected to the fusion valley (full red circles) through the 
barrier "V1$\to$fus" whose height is $\simeq$3~MeV. Then, it begins to 
decrease and goes away at a larger deformation Q$_{20} \simeq 202$~b. The 
exit point of the principal valley V1 is obtained for Q$_{20} \simeq 
206$~b and Q$_{40} \simeq 255$~b$^{2}$.

In the $^{216}$Th isotope, one obtains  
for the first time a valley V2 which is lower in energy than the 
principal valley V1 as can be observed in Fig. \ref{collQ404} (d), 
when using the D1ST2a interaction. The 
exit point of the valley V2 is still strongly compatible with a compact 
fission mode, with Q$_{20} \simeq 162$~b and Q$_{40} \simeq 150 {\rm 
b}^2$. For the principal valley V1, the exit point has Q$_{20} \simeq 
190$~b and Q$_{40} \simeq 225$~b$^2$. In the calculations with the 
D1S interaction a unique valley V1, corresponding to a well-elongated 
fission mode, is obtained. This valley ends at Q$_{20} \simeq 198$~b and Q$_{40} \simeq 240$~b$^2$.

Looking at the barrier heights in Fig. \ref{barrierQ40a} (d), one 
observes that all the barriers, namely "V1$\to$fus", "V2$\to$fus" and 
"V1$\to$V2", start with a lower energy than the equivalent ones 
obtained in the previous isotopes with either D1ST2a or D1S Gogny 
interaction. In particular, at Q$_{20} \simeq 130$~b, the 
"V2$\to$fus" barrier (full blue triangles) is higher in energy than the 
"V1$\to$fus" one. This is consistent with the fact that the principal 
valley V1 is located above the V2 one. This observation remains true up 
to Q$_{20} \simeq 158$~b where the valley V2 disappears. Here, the 
"V1$\to$fus" barrier height starts to increase again and reaches 
$\simeq 2$~MeV at Q$_{20} \simeq 172$~b. Then, it decreases and 
fades away at Q$_{20} \simeq 190$~b.

It is also interesting to note that the barrier "V1$\to$V2" (green 
stars) is very low at the beginning of the curve, around $\simeq 1$~MeV. 
Then, it decreases with the quadrupole deformation up to Q$_{20} \simeq 
152$~b, where it reaches $\simeq$ $300$~keV. It increases again up to 
Q$_{20} \simeq 158$~b where reaches the  $\simeq 800$~keV 
height and disappears at the end of the valley V2. Finally, for the D1S 
interaction, the behavior of the "V1$\to$fus" barrier is similar to 
the one obtained in the $^{222}$Th isotope, but with a little shorter 
extension before disappearing.

All the characteristics of the exit points, in terms of values of the 
collective variables Q$_{20}$, Q$_{30}$ and Q$_{40}$, and associated 
with the symmetric and the asymmetric valleys, are given in Tables 
\ref{SYMEXITD1S}, \ref{SYMEXITD1ST2a}, \ref{ASYMEXITD1S}, and 
\ref{ASYMEXITD1ST2a}, for both the D1S and the D1ST2a interactions.

Preliminaries calculations in neutron deficient uranium and radium 
isotopes seem to display also a second symmetric valley corresponding 
to a compact fission mode.

\begin{table}[!] \centering
 \begin{tabular}{|c|c|c|c|}
\hline
  Nucleus     &   Valley &   Q$_{20}$ (b)     &  Q$_{40}$  (b$^{2}$)         \\ 
\hline	
\hline
$^{230}$Th     &    V1     &  210               &       255                     \\
\hline	
$^{226}$Th     &    V1     &  208               &       255                   \\
\hline	
$^{222}$Th     &    V1     &  206               &       255                 \\
\hline	
$^{216}$Th     &    V1     &  198               &       240                  \\
\hline	
\hline
\multicolumn{4}{|c|}{D1S, Symmetric path} \\
\hline
 \end{tabular}
\caption{Values of the Q$_{20}$ (expressed in b) and Q$_{40}$ (expressed in b$^{2}$) collective variables of the exit points 
in the $^{230}$Th, $^{226}$Th, $^{222}$Th and $^{216}$Th isotopes. Calculation have been done for the symmetric path with the 
D1S Gogny force.}
\label{SYMEXITD1S} 
\end{table}

\begin{table}[!] \centering
 \begin{tabular}{|c|c|c|c|c|}
\hline
  Nucleus      &  Valley   &    Q$_{20}$ (b)     &  Q$_{40}$  (b$^{2}$)        \\ 
\hline
\hline	
$^{230}$Th     &    V1     &     230             &       315                    \\
\hline
$^{226}$Th     &    V1     &     222             &       295                   \\
\hline	
               &    V2     &     {\bf 166}       &       150                     \\
\hline	
$^{222}$Th     &    V1     &     208             &       265                    \\
\hline
               &    V2     &     {\bf 166}       &       150                     \\
\hline	
$^{216}$Th     &    V1     &     190             &       225                  \\
\hline	
               &    V2     &     {\bf 162}       &       150                   \\
\hline	
\hline
\multicolumn{4}{|c|}{D1ST2a, Symmetric path} \\
\hline
 \end{tabular}
\caption{Same as Table \ref{SYMEXITD1S} but for the D1ST2a Gogny interaction.}
\label{SYMEXITD1ST2a} 
\end{table}

\begin{table}[!] \centering
 \begin{tabular}{|c|c|c|c|c|c|}
\hline
  Nucleus     &   Valley &   Q$_{20}$ (b) &   Q$_{30}$ (b$^{3/2}$)    &  Q$_{40}$  (b$^{2}$)          \\ 
\hline	
\hline
$^{230}$Th     &    V1     &  136     &     28     &        80                     \\
\hline	
$^{226}$Th     &    V1     &  140     &     40     &        94                   \\
\hline	
$^{222}$Th     &    V1     &  172     &     69     &       159                  \\
\hline	
$^{216}$Th     &    V1     &  150     &     52     &       116                  \\
\hline	
\hline
\multicolumn{5}{|c|}{D1S, Asymmetric path} \\
\hline
 \end{tabular}
\caption{Same as Table \ref{SYMEXITD1S} but for the asymmetric valley.}
\label{ASYMEXITD1S} 
\end{table}

\begin{table}[!] \centering
 \begin{tabular}{|c|c|c|c|c|c|}
\hline
  Nucleus     &   Valley &   Q$_{20}$ (b) &   Q$_{30}$ (b$^{3/2}$)    &  Q$_{40}$  (b$^{2}$)         \\ 
\hline	
\hline
$^{230}$Th     &    V1     &  136     &     33     &        85                    \\
\hline	
$^{226}$Th     &    V1     &  134     &     39     &        87                    \\
\hline	
$^{222}$Th     &    V1     &  162     &     64     &       142                    \\
\hline	
$^{216}$Th     &    V1     &  150     &     51     &       116                   \\
\hline	
\hline
\multicolumn{5}{|c|}{D1ST2a, Asymmetric path} \\
\hline
 \end{tabular}
\caption{Same as Table \ref{SYMEXITD1ST2a} but for the asymmetric valley.}
\label{ASYMEXITD1ST2a} 
\end{table}

\begin{figure}[!] \centering
\begin{tabular}{cc}
\includegraphics[width=8.5cm]{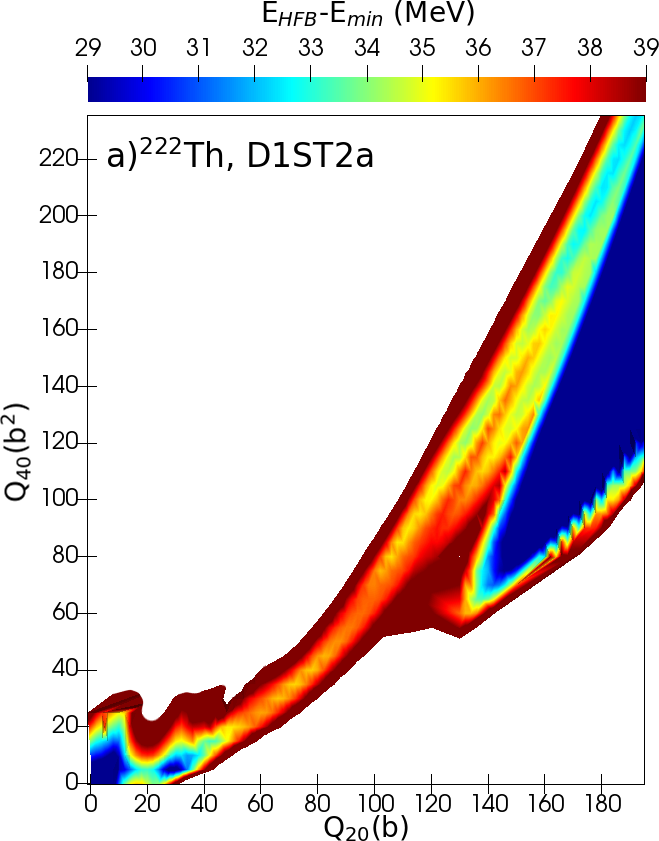}  \vspace{0.3cm} \\
\includegraphics[width=8.5cm]{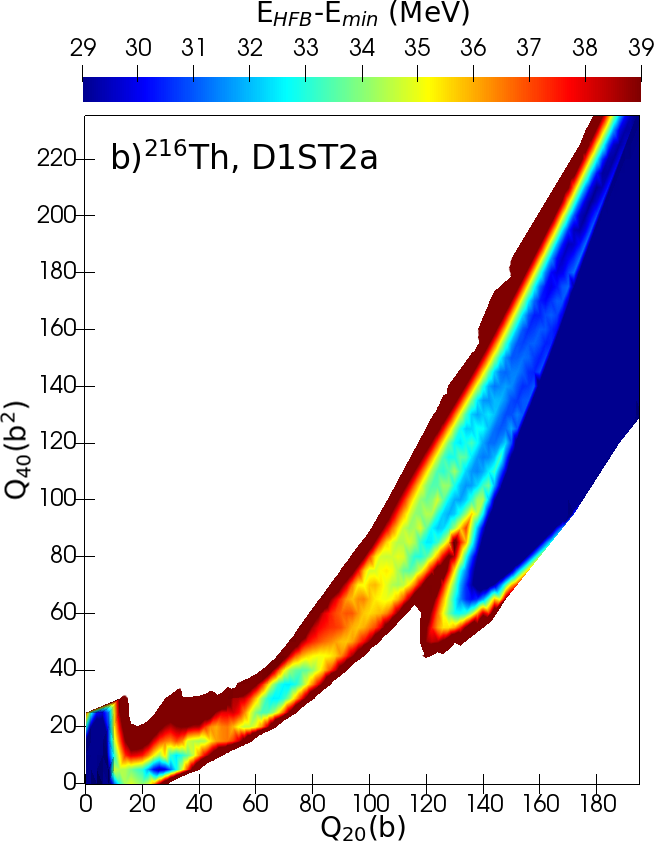}
\end{tabular}
\caption{Creation of the two valleys using Q$_{20}$ and Q$_{40}$ as collective variables along the symmetric path,
in (a) $^{222}$Th and (b) $^{216}$Th with the D1ST2a interaction.
Energies are expressed in MeV.} 
\label{barrierQ40b}
\end{figure}

To finish this section let us say a few words about the origin of the 
valleys V2 which are well pronounced in the $^{222}$Th and $^{216}$Th 
isotopes. In Fig. \ref{barrierQ40b}, a 2D representation of the 
symmetric $\{$Q$_{20}$, Q$_{40}$$\}$ PES's calculated with the D1ST2a 
interaction for both isotopes is displayed. Here, results include 
smaller values of Q$_{20}$, namely Q$_{20}= 0$~b and beyond. 
Besides, we have kept the fusion valley which appears in dark blue 
on the right side of the fission valleys.

In Fig. \ref{barrierQ40b} (a) we observe the existence of a plateau in 
$^{222}$Th which extends between Q$_{20} \simeq 90$ b and $\simeq$100 
b. Just after the plateau, one sees the nascence of the valley V1. 
Concerning the valley V2, it appears at a much larger elongation, 
around Q$_{20} \simeq 128$~b. Invoking only the topology of the 
PES and this difference in quadrupole deformation, we can argue that, 
in the adiabatic approximation, the flux of the wave function will feed 
directly and largely the valley V1 in the region Q$_{20} \in [100~{\rm 
b}, 130~{\rm b}]$. In addition, using the findings of the (Q$_{20}$, 
Q$_{40}$) dynamical study proposed by J.F. Berger and collaborators in 
the context of cold fission for the $^{240}$Pu nucleus \cite{Gogny5}, 
the valley V2 may be fed partly by the wave function through the 
excitation of transverse modes for larger elongations. The height of 
the barrier "V1$\to$V2" discussed previously (see Fig. 
\ref{barrierQ40a} (c), green stars) is fully compatible with such a 
process. In that context, the symmetric fission is understood as a 
mixing of a compact and the super long modes whose weight can be 
determined by a dynamical treatment.

In $^{216}$Th (Fig. \ref{barrierQ40b} (b)), the pattern is rather 
different. Indeed, the V2 valley appears first around Q$_{20} \simeq 
110$~b and it is the lowest in energy. From the plateau in energy which 
exists between $88$~b and $100$~b and the lowest energy path which is 
located on the side of the lowest Q$_{40}$ value, one concludes that, 
this time, the valley V2 will be the one preferentially fed by the time 
evolution of the collective wave function. As in $^{222}$Th and 
considering the heights of the barriers "V2$\to$fus" and "V1$\to$V2" 
(see Fig. \ref{barrierQ40a} (d), full blue triangles and green stars, 
respectively), one predicts an exchange between the valleys V2 and V1 
through transverse modes and the manifestation of both the compact and 
super long symmetric modes.

\subsection{Interplay between tensor force, deformation and pairing correlations}\label{resultsC}

We now turn our attention to the mechanism responsible for the existence 
of the new valley V2, which is interpreted as the experimentally 
observed new symmetric compact mode. The present analysis has been done 
by inspecting the different contributions to the total HFB energy. More 
precisely, we have separated the HFB binding energy in two (three) 
contributions in the case of the D1S (D1ST2a) interaction in such a way 
that:
\begin{equation}
\begin{array}{lcl}
{\rm E}_{\rm HFB}^{\rm D1S} &=& {\rm E}_{\rm MF}^{\rm D1S}+ {\rm E}_{\rm pair}^{\rm D1S} \\
{\rm E}_{\rm HFB}^{\rm D1ST2a} &=& {\rm E}_{\rm MF}^{\rm D1ST2a}+ {\rm E}_{\rm pair}^{\rm D1ST2a}+{\rm E}_{\rm TS}
\end{array}
\end{equation} 
where ${\rm E}_{\rm MF}$ is the mean-field energy, not including the 
tensor contribution in the D1ST2a case. The particle-particle energy 
${\rm E}_{\rm pair}=\frac{1}{2}\mathrm{Tr}\, (\Delta \kappa)$ is usually 
referred to as the pairing energy and is proportional to the amount of 
pairing correlations in the system. It should not be confused with the 
real pairing correlation energy given by the difference between the HFB 
and HF energies. Finally, the tensor energy ${\rm E}_{\rm TS}$ is the 
contribution of the tensor term to the HFB energy and therefore it is 
zero in the D1S case.

\begin{figure*} \centering
\includegraphics[width=18cm]{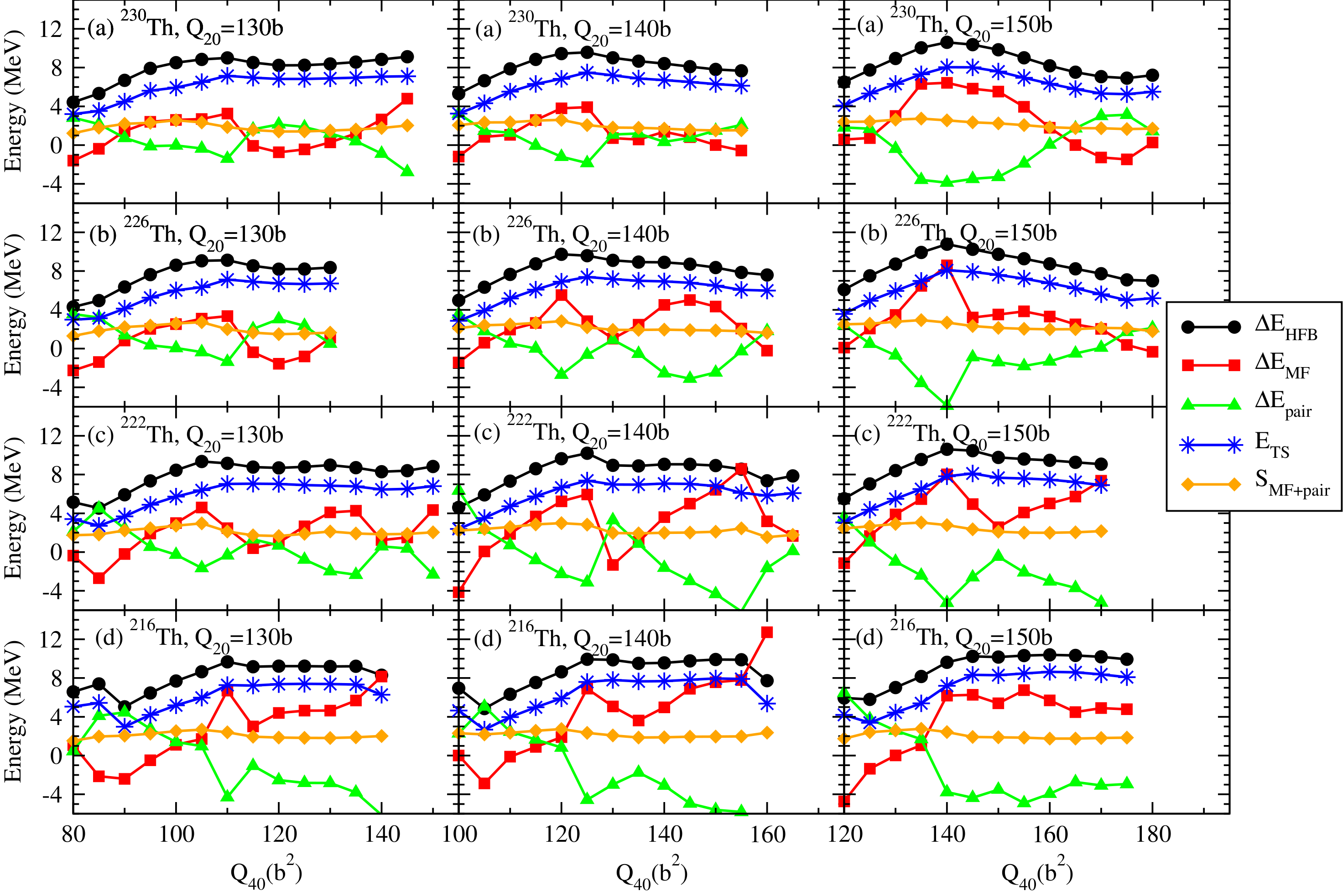} 
\caption{Evolution of $\Delta {\rm E}_{{\rm HFB}}$, $\Delta {\rm E}_{{\rm MF}}$,
$\Delta {\rm E}_{{\rm pair}}$ and ${\rm E}_{{\rm TS}}$
(see text for explanation) as a function of Q$_{40}$ (in b$^2$) for (a) $^{230}$Th (b) $^{226}$Th (c) $^{222}$Th (d) $^{216}$Th.
Calculations have been done at Q$_{20}=130$~b (left column),$140$~b (central column) and $150$~b (right column). Energies are expressed in MeV.} 
\label{decomp1}
\end{figure*}

In Fig. \ref{decomp1}, we display the energy differences:
\begin{equation}
\begin{array}{lcll}
\Delta {\rm E}_{{\rm HFB}} & = & {\rm E}_{{\rm HFB}}^{{\rm D1ST2a}} - 
{\rm E}_{{\rm HFB}}^{{\rm D1S}} &~~\mbox{(full black circles)}, \nonumber \\ 
\Delta {\rm E}_{{\rm MF}}&=& {\rm E}_{{\rm MF}}^{{\rm D1ST2a}} - 
{\rm E}_{{\rm MF}}^{{\rm D1S}} &~~\mbox{(full red squares)}, \nonumber \\
\Delta {\rm E}_{{\rm pair}}&=& {\rm E}_{{\rm pair}}^{{\rm D1ST2a}} - 
{\rm E}_{{\rm pair}}^{{\rm D1S}} &~~\mbox{(full green triangles)} \nonumber
\end{array}
\end{equation} 
as well as E$_{TS}$ (blue stars), as a function of the hexadecapole 
moment Q$_{40}$ expressed in b$^{2}$. We have also depicted the 
accumulated sum S$_{{\rm MF+pair}}= \Delta {\rm E}_{{\rm MF}}+ \Delta 
{\rm E}_{{\rm pair}}$ (orange full diamonds). The calculations have 
been performed with the additional constraint Q$_{20}=130$~b (left 
column), $140$~b (central column) and $150$~b (right column), for (a) 
$^{230}$Th (b) $^{226}$Th (c) $^{222}$Th and (d) $^{216}$Th. 

The most streaking feature observed in all the panels is the similar 
behaviour of $\Delta {\rm E}_{{\rm HFB}}$, whatever the isotope and the 
deformation Q$_{20}$. Starting from the barrier which separates the 
fusion and the fission valleys at the smallest Q$_{40}$ values, one 
first observes a linear increase of $\Delta {\rm E}_{{\rm HFB}}$. The 
positive sign of this variation indicates that the D1ST2a interaction 
produces less binding energy than the D1S one. A maximum is obtained at 
a Q$_{40}$ value which corresponds to the ridge between the new valley 
V2 (when it exists) and the main valley V1. One notes that, even if the 
valley V2 is not apparent, as it is the case in $^{230}$Th, a maximum 
for E$_{{\rm TS}}$ is also obtained in the same Q$_{40}$ region. In 
addition, the intensity of the phenomenon is nearly the same for all 
the isotopes. It starts around $3-4$~MeV at the fusion-fission barrier 
with a variation of $3-4$~MeV. After the maximum reached by E$_{{\rm 
TS}}$, $\Delta {\rm E}_{{\rm HFB}}$ decreases or stabilizes in several 
cases.

\begin{figure}[!] \centering
\begin{tabular}{c}
\includegraphics[width=8.5cm]{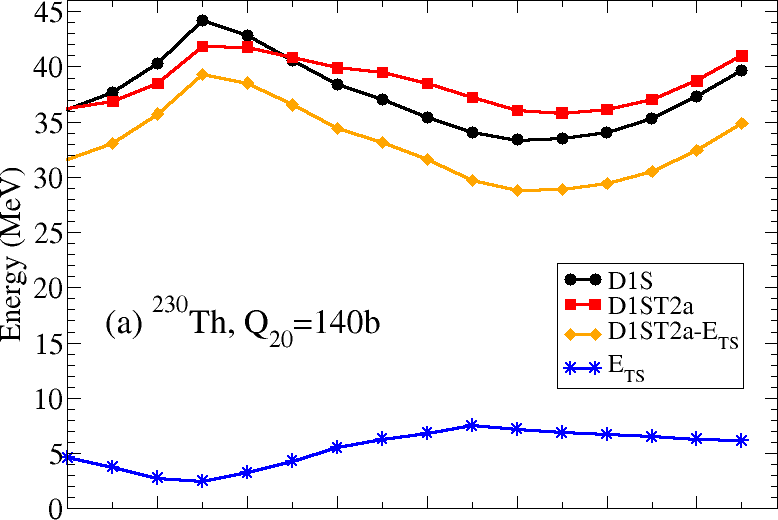} \\ 
\includegraphics[width=8.5cm]{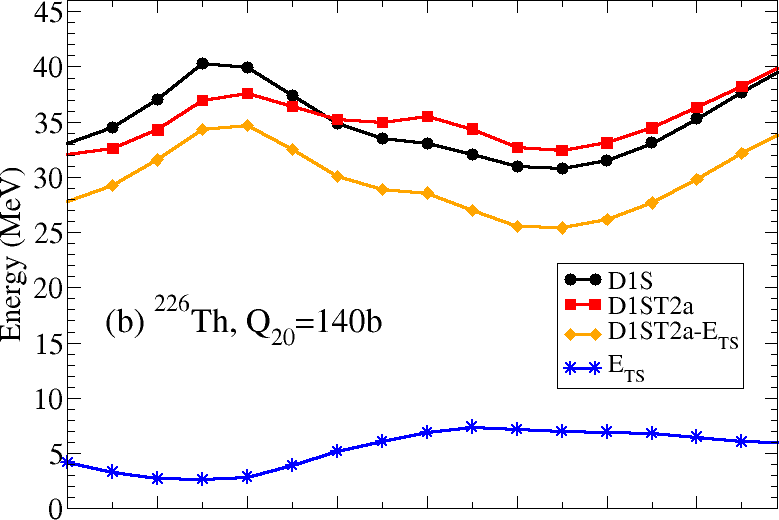} \\
\includegraphics[width=8.5cm]{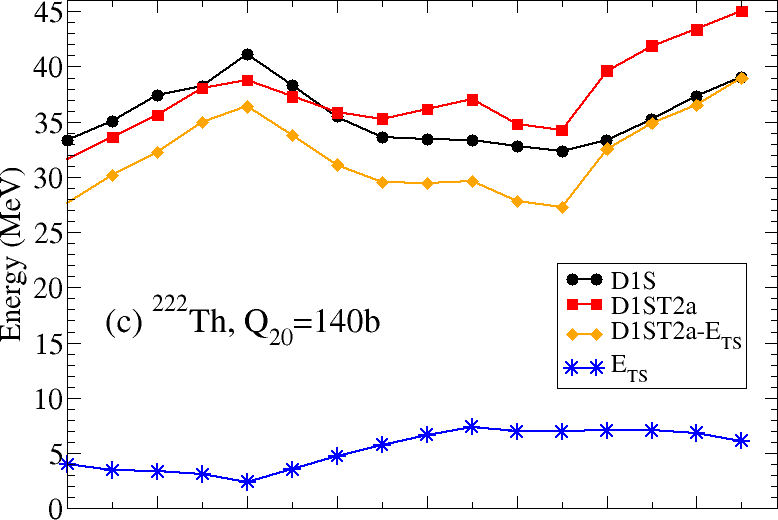} \\ 
\includegraphics[width=8.5cm]{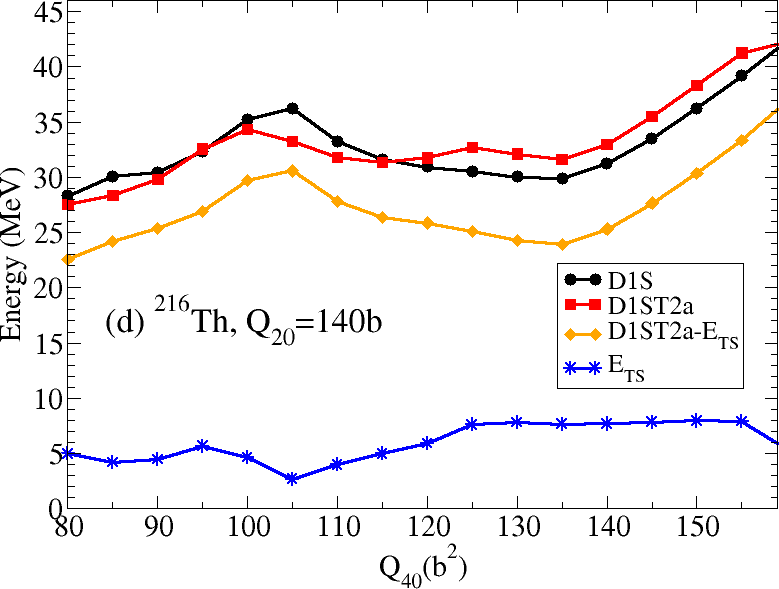} 
\end{tabular}
\caption{Evolution of E$_{{\rm HFB}}$-E$_{{\rm HFB}}({\rm g.s})$ as a 
function of Q$_{40}$ calculated with D1S (black full circles) and 
D1ST2a (red full squares). Moreover, we have drawn the curve 
corresponding to adding the tensor energy E$_{TS}$ to the D1S energy 
(orange full diamonds). For comparison, the tensor contribution 
E$_{\mbox{TS}}$ (blue stars) is also depicted. Energies are expressed in 
MeV.} 
\label{truc}
\end{figure}
\begin{figure} \centering
\begin{tabular}{c}
\includegraphics[width=8.3cm]{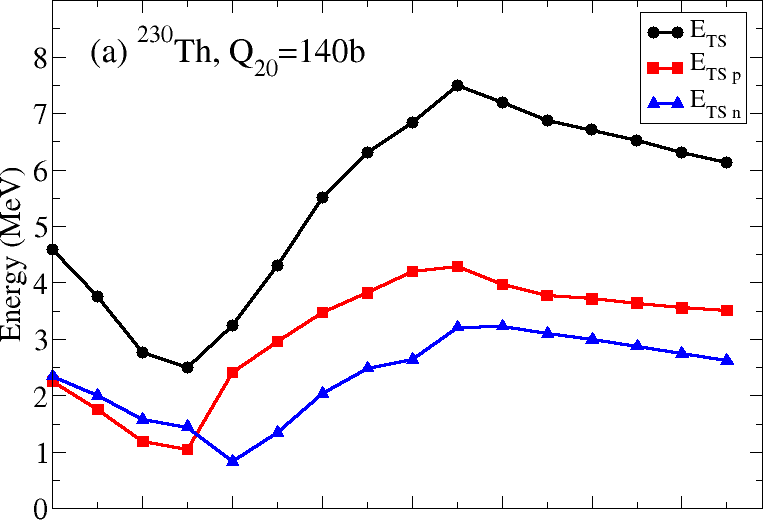} \\ 
\includegraphics[width=8.3cm]{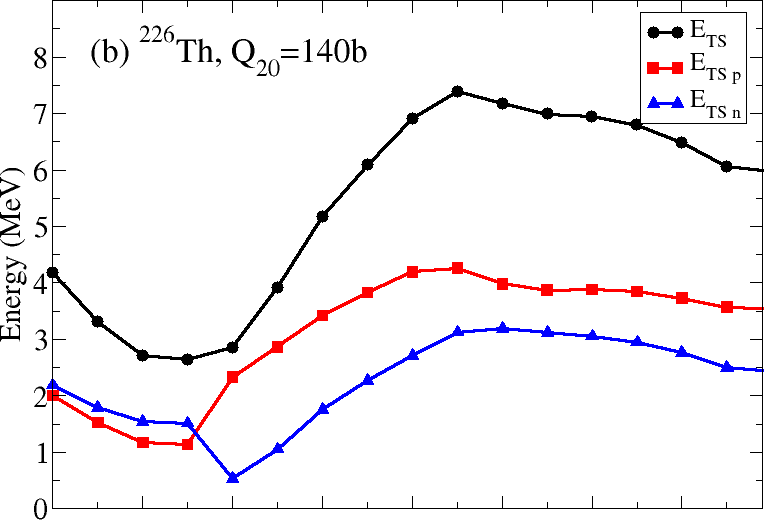} \\ 
\includegraphics[width=8.3cm]{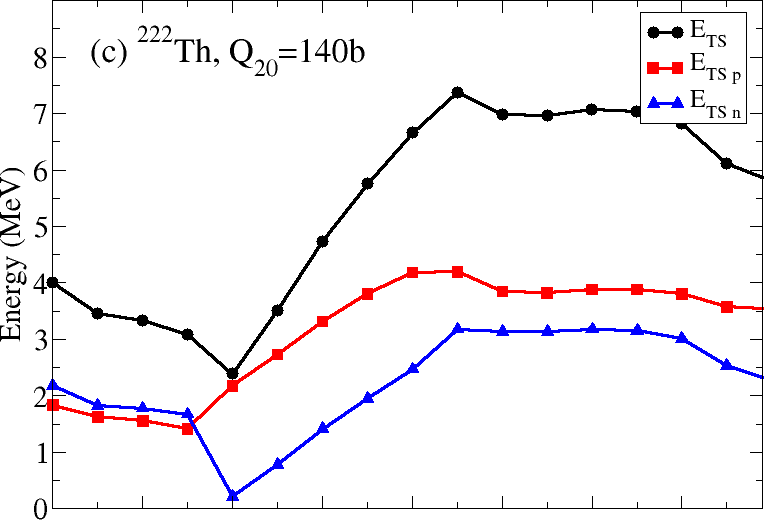} \\ 
\includegraphics[width=8.3cm]{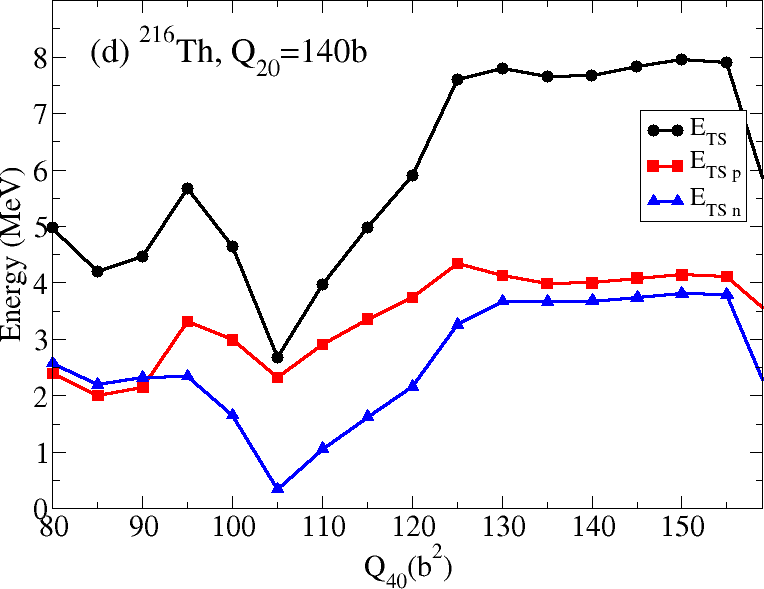} 
\end{tabular}
\caption{Evolution of the total tensor energy E$_{TS}$ (black full 
circles), its proton component E$_{\rm TS p}$ (red full squares) and 
its neutron component E$_{\rm TS n}$ (blue full triangles) as a 
function of Q$_{40}$ (expressed in b$^2$) at Q$_{20}=130$~b. See the 
text for details. Calculations have been done for (a) $^{230}$Th (b) 
$^{226}$Th (c) $^{222}$Th (d) $^{216}$Th. Energies are expressed in 
MeV.} 
\label{tensor1}
\end{figure}

The general trend obtained for $\Delta {\rm E}_{{\rm HFB}}$ seems to be 
strongly correlated with the E$_{\rm TS}$ contribution and this 
represents a first strong hint of a tensor effect. In order to better 
isolate this effect, we have plotted in Fig. \ref{truc} the evolution 
of E$_{{\rm HFB}}$-E$_{{\rm HFB}}({\rm g.s})$ as a function of 
Q$_{40}$, calculated with D1S (full black circles) and D1ST2a (full red 
squares). In the same figure, the curve corresponding to the results 
obtained with the D1ST2a force but subtracting the tensor energy 
E$_{\rm TS}$ (full orange diamonds) is also shown. To facilitate the 
interpretation of the results, the quantity E$_{{\rm TS}}$ (blue stars) 
is also drawn. As an illustration, calculations are shown for the four 
isotopes with the constraint Q$_{20}={\rm 140}~b$. The similarity of 
the "D1S" and "D1ST2a-E$_{{\rm TS}}$" curves leads to the conclusion 
that the birth of the new valley V2 is due to the increase of E$_{\rm 
TS}$ with the hexadecapole moment up to a certain value of Q$_{40}$ 
which is coupled to the decreasing slope of the MF plus pairing 
contributions obtained in this region. This effect is not sufficient in 
$^{230}$Th to create a new valley V2. However, the slope of the curve 
is softened by the tensor contribution (see the curves with red square 
and orange diamonds). 
The preservation of the valley V1 for larger values of Q$_{40}$ is due 
to the decrease or stabilization of E$_{{\rm TS}}$. One notes that in 
$^{222}$Th there is a local effect around Q$_{40}\simeq 135{\rm b}^2$ 
for the "D1ST2a-E$_{{\rm TS}}$" curve which leads to a more pronounced 
minimum than for the "D1S" curve. To end with the effect associated 
with the E$_{\rm TS}$ contribution, we have displayed in Fig. 
\ref{tensor1} the evolution of the proton E$_{\rm TS p}$ (full red 
squares) and the neutron  E$_{\rm TS n}$ (full blue triangles) 
component of the total tensor energy E$_{\rm TS}$ (black full circles) 
as a function of Q$_{40}$, for all the considered isotopes. We observe 
that in the region of the new valley, both proton and neutron 
contributions increase with the proton one dominating over the 
neutrons. For larger values of Q$_{40}$, they show a rather constant 
behavior with similar contributions in the two cases.

The mean-field E$_{{\rm MF}}$ and the pairing E$_{{\rm pair}}$ energy 
contributions depicted in Fig. \ref{decomp1} suffer from strong 
variations when the tensor term is added to the D1S interaction. They 
vary out of phase with changes of sign for both contributions. When the 
mean-field is less bound with the D1ST2a interaction ($\Delta {\rm 
E}_{{\rm MF}} \ge 0$), the pairing correlations increase and {\it vice 
versa}. 
The sum of the two quantities has a positive value as can be 
seen in the curve S$_{{\rm MF+pair}}$. 
Besides, the behavior of the 
quantity is found to be rather constant (the variations are within less 
than 1 MeV). This last result confirms the role played by the tensor 
energy E$_{{\rm TS}}$ in the creation of the new valley, which was 
discussed previously.

\begin{figure} \centering
\begin{tabular}{c}
\includegraphics[width=8.5cm]{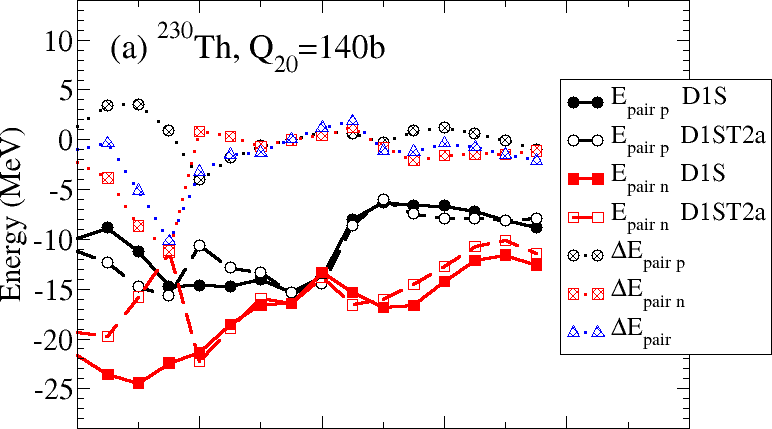} \\ 
\includegraphics[width=8.5cm]{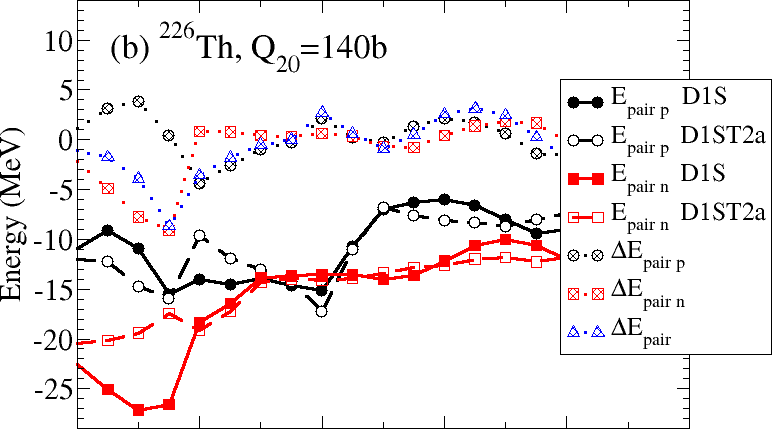} \\
\includegraphics[width=8.5cm]{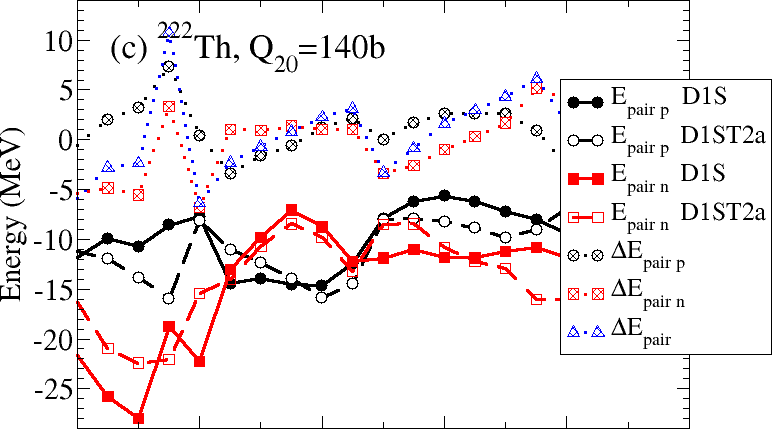} \\ 
\includegraphics[width=8.5cm]{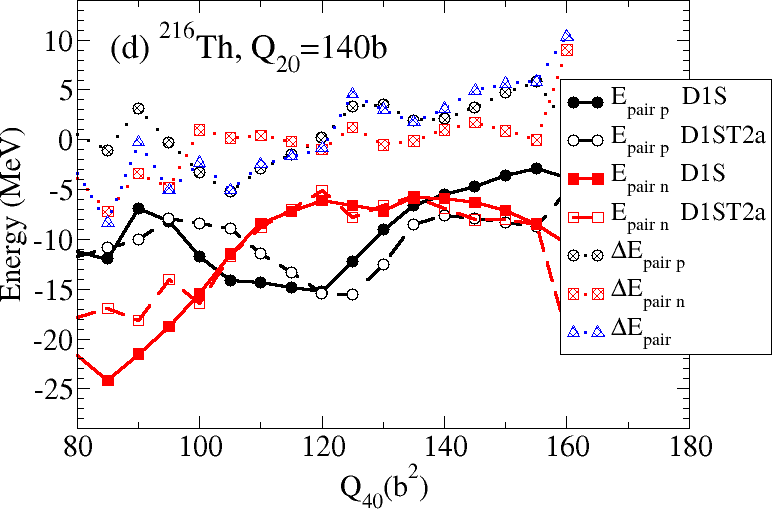} 
\end{tabular}
\caption{Evolution of proton and neutron pairing components for both 
the D1S (black full circles and red full squares, respectively) and 
D1ST2a (black empty circles and red empty squares,respectively) 
interactions. The proton $\Delta \mbox{E}_{\mbox{\rm pair p}}$ and 
neutron $\Delta \mbox{E}_{\mbox{\rm pair n}}$ differences between both 
interactions (black dashed circles and red dashed squares are also 
indicated. The total difference $\Delta \mbox{E}_{\rm pair}$ is shown 
in blue dashed triangles. Calculations have been done for (a) 
$^{230}$Th, (b) $^{226}$Th, (c) $^{222}$Th and (d) $^{216}$Th at 
Q$_{20}= 140$~b. Energies are expressed in MeV.} 
\label{Pairing1}
\end{figure}

We would like to end this part by discussing the pairing contribution. 
Indeed, even though from a total energy perspective the role of the 
pairing seems to be washed out by the mean field contribution, many 
observables are sensitive to these correlations, as for example the 
collective masses which are crucial for the dynamical propagation. In 
Fig. \ref{Pairing1}, we report the evolution of the proton and neutron 
pairing components for both the D1S (full black circles and full red 
squares, respectively) and D1ST2a (empty black circles and empty red 
squares, respectively) interactions. Moreover, the proton $\Delta 
\mbox{E}_{\mbox{\rm pair p}}$ and neutron $\Delta \mbox{E}_{\mbox{\rm 
pair n}}$ differences between both interactions (dashed black circles 
and dashed red squares, respectively) are also drawn. Finally, the 
total difference $\Delta \mbox{E}_{\rm pair}$ in shown with dashed blue 
triangles. Calculations have been done for (a) $^{230}$Th, (b) 
$^{226}$Th, (c) $^{222}$Th and (d) $^{216}$Th. As an example, the 
figure shows results for Q$_{20} = 140$ b. One observes a similar trend 
for all the isotopes. From the fusion-fission ridge and the V2-V1 ridge 
(A area), there is an increase of $\Delta \mbox{E}_{\rm pair}$. Then, a 
decrease is obtained between the V2-V1 ridge and the bottom of the 
valley V1 (B area) and finally a new increase for larger value of 
Q$_{40}$ (C area) manifests. Looking at the proton and neutron 
decomposition, one sees that  the variation $\Delta \mbox{E}_{\rm 
pair}$ in the A area is mainly due to the proton pairing variation. The 
neutron one is nearly constant and close to zero. The proton variation 
changes sign whereas the neutron one is positive. In the B area, both 
proton and neutron variations decrease and participate in the total 
decrease which is found moderate. In the C area, the behavior of 
$\Delta \mbox{E}_{\rm pair}$ in terms of proton and neutron components 
depends on the nucleus.

The neutron pairing energy along the isotopic  chain is found to be 
very similar for both interactions in the region of the valley V2, with 
a value that changes a lot from  isotope to isotope. Indeed, one 
observes a strong decrease from the heaviest to the lightest thorium 
isotopes. On the other hand, some variations appear in the valley V1. 
Concerning the proton pairing energy, differences in both the A and B 
areas are observed. In general, the proton pairing is larger in the A 
area and smaller in the B area with the D1S interaction 

\subsection{Distribution of the available energy at scission and neutron multiplicity}\label{resultsD}

In this section we discuss the way the available energy of the 
fissioning system is distributed among the various physical components 
at scission. The available energy is defined as the difference between 
the total energy of the fissioning nucleus E$_{\rm tot}$ and the sum of 
the ground state energy of fragments E$^{\rm g.s.}_{\rm frag}$. At 
scission, the available energy goes into two contributions: the total 
kinetic energy (TKE) and the total excitation energy (TXE), 
\begin{equation}
\displaystyle \mathrm{E}_{\rm tot} - \mathrm{E}^{\rm ~g.s.}_{\rm frag} = \mathrm{TKE} + \mathrm{TXE} 
\end{equation}  
The TKE takes most of the available energy and it is dominated by the 
Coulomb repulsion E$_{\rm Coul}$ energy between the fragments. The 
remaining part is known as the pre-kinetic energy E$_{\rm prek}$. 
Concerning the TXE, it is the sum of the deformation energy E$_{\rm 
def}$ of the fragments  and their intrinsic excitation energy E$_{\rm 
intr}$. Thus,
\begin{equation}
\displaystyle \mathrm{E}_{\rm tot} = \mathrm{\rm E}_{\rm Coul} + \mathrm{\rm E}_{\rm prek} + \mathrm{\rm E}_{\rm def} + \mathrm{\rm E}_{\rm intr} + 
\mathrm{\rm E}^{\rm ~g.s.}_{\rm frag} 
\label{DisE}
\end{equation}  
In the present study, as we discuss low energy fission, we choose the 
total energy E$_{\rm tot}$ as the HFB energy obtained at the saddle 
point. Here, the scission point is defined when a sudden drop of the 
density between pre-fragments occurs. All the quantities involved in 
the energy distribution are evaluated in the first point of the PES 
mesh when the fragments appear.

The Coulomb energy is calculated at the scission point using the simple 
Coulomb formula: 
\begin{equation}
 \mathrm{E}_{\rm Coul} = \frac{\mathrm{Z}_{1} \mathrm{Z}_{2} e^{2}}{\rm d_{\rm ch}}
\end{equation}  
where Z$_{1}$ and Z$_{2}$ define the charge of the two fragments and 
$\rm d_{\rm ch}$ is the distance between the center of mass of the 
charge distributions of the fragments at the exit point. Coulomb 
energies obtained for the symmetric valleys and for the four Thorium 
isotopes are reported in Table \ref{TKED1S} (D1S) and Table 
\ref{TKED1ST2a} (D1ST2a). The Coulomb energy is almost isotope 
independent for the results with D1S due to the fact that the scission 
point are in the same quadrupole moment region (see Table 
\ref{SYMEXITD1S}). The same conclusion apply for the D1ST2a interaction 
in valley V2 (see Table \ref{SYMEXITD1ST2a}). However, in the  V1 
valley and with the D1ST2a interaction we find that the heavier the 
Thorium isotope, the larger the elongation for the scission point. As a 
result, the distance $\rm d_{ch}$ is larger for heavy isotopes and the 
Coulomb energy is smaller. Coulomb energies from valley V2 are always 
bigger than the ones in valley V1 since their exit point occur at a 
smaller deformation.

The energies of the fragments at scission E$_{\rm frag}$ and the 
corresponding energies when the two fragments are well separated 
E$^{\rm ~g.s.}_{\rm frag}$ are obtained by means of HFB calculations. 
The deformation energy E$_{\rm def}$ is the differences between these 
two energies. Quadrupole and octupole moments from the fragments at 
scission are used as constraints to get the fragment HFB energy E$_{\rm 
frag}$. The symmetric fission in the $^{230}$Th, $^{226}$Th, $^{222}$Th 
and $^{216}$Th isotopes leads to $^{115}$Rh, $^{113}$Rh, $^{111}$Rh and 
$^{108}$Rh fragments, respectively. In this work, the equal filling 
approximation has been used to calculate both the ground state and the 
deformed Rhodium isotopes using the same kind of methodology as the one 
reported in \cite{RobPRC12} for odd and odd-odd nuclei. Fragment 
deformation energies E$_{\rm def}$ are depicted in Table~\ref{TKED1S} 
and Table~\ref{TKED1ST2a}. As expected, in both calculations with D1S 
and D1ST2a, the deformation energy is bigger for more elongated 
fission. The most striking feature is that D1ST2a provides more 
deformation energy than D1S. This is expected when the scission point 
elongation is bigger for D1ST2a than D1S such as for $^{230}$Th and 
$^{226}$Th but it remains true when the elongation is about the same 
($^{222}$Th) or is smaller ($^{216}$Th). Besides, for D1ST2a, 
deformation energies from valley V2 are significantly smaller than the 
ones in valley V1 by a factor 4.

Once Coulomb and fragment energies are calculated, Eq.~(\ref{DisE}) 
provides the quantity $\mathrm{E}_{\rm prek}+\mathrm{E}_{\rm intr}$ 
that are discussed in the following. Since quasi-particle excitation is 
not considered in this work to build PES, a microscopic evaluation of 
the part of the total available energy which is converted to intrinsic 
excitation is out of the scope of this work. Such a task using the 
Generator Coordinate Method framework would require the use of a non 
adiabatic model such as in \cite{BerPRC11}. We thus introduce three 
different scenarios about the way the energy is shared between the 
pre-kinetic and intrinsic energies and thus between the TKE and the 
TXE. Scenario 1 is defined as the one in which all the available energy 
$\mathrm{E}_{\rm prek}+\mathrm{E}_{\rm intr}$ goes to the pre-kinetic 
energy: $\mathrm{E}_{\rm intr}=0$~MeV. In the second scenario, the 
intrinsic energy is chosen according to the empirical formula 
$\mathrm{E}_{\rm intr}=35\%\mathrm{TXE}$ used in \cite{CaaPLB17, 
RejNPA00, SchNDS16}. Contrary to scenario 1, in scenario 3, all the 
available energy goes to the intrinsic energy exclusively: 
$\mathrm{E}_{\rm prek}=0$~MeV. Even if they may not be realistic in 
some cases, scenarios 1 and 3 provide boundaries for the quantities 
under consideration. For the three scenarios, the TKE and TXE has been 
obtained and given in Tables~\ref{TKED1S} and \ref{TKED1ST2a}. For D1S, 
scenarios 1 and 2 look very similar: the TKE is stable for the three 
heaviest isotopes and becomes significantly bigger for $^{216}$Th. In a 
consistent way, the TXE is almost constant and decreases for 
$^{216}$Th. In scenario 3 the TKE and TXE remain almost constant along 
the isotopic chain. For the D1ST2a interaction, the TKE is always 
driven by the Coulomb energy: for all the scenarios it increases with 
the isotope exoticism in valley V1 and is stable in valley V2. The 
three TXE are driven by the deformation energy which decreases with the 
mass number. Since the deformation energy is small in valley V2, 
scenario 2 is much closer to scenario 1 than scenario 3 for both TXE 
and TKE.

The TKE have been measured in \cite{BocNPA08} for $^{226}$Th. The super 
long symmetric mode gives $\mathrm{TKE} \simeq 160$~MeV close to 
scenario 3 for D1S ($157.2$~MeV) and scenario 2, valley V1 
($161.0$~MeV) for D1ST2a. This latter valley is the one energetically  
preferred for this isotope. In \cite{Schmidt} the overall mean TKE in 
the Thorium chain from $^{229}$Th to $^{221}$Th is given. It is stable 
along the symmetric / asymmetric transition with 
$\langle\mathrm{TKE}\rangle=167.7 \pm 3.4$ for $^{226}$Th and 
$\langle\mathrm{TKE}\rangle=166.9 \pm 3.3$ for $^{222}$Th. A comparison 
with the mean TKE will be possible in the future by solving the TDGCM 
equations with the static PES.

\begin{table*} \centering
 \begin{tabular}{|c|c|c|c|c|c|c|c|c|c|c|c|c|}
\hline
 Nucleus    &  Valley   & ~~TKE$_{1}$ ~~& ~~TKE$_{2}$ ~~&~~ TKE$_{3}$~~ & ~~E$_{\rm coul}$  ~~ &~~ TXE$_{1}$~~ & ~~TXE$_{2}$ ~~&~~TXE$_{3}$~~& ~~E$_{\rm def}$ ~~& $~~ \nu_{1}$ ~~&~~ $ \nu_{2} $~~ &~~ $ \nu_{3} $~~    \\ 
\hline	
\hline
$^{230}$Th     &    V1        &   178.7 &  170.4  & 157.8  &  157.8   &  15.4   &  23.6   & 36.3  & 15.4   &    $\sim$1  &        1     &  2 \\
\hline
$^{226}$Th     &    V1        &   178.0 &  170.3  & 157.2  &  157.2  &  14.2   &   21.9   & 35.0  & 14.2   &    0        &        1     &   2  \\
\hline	
$^{222}$Th     &    V1        &   177.0 &  169.1  & 156.5  &  156.5  &  14.8   &   22.8   & 35.3  & 14.8   &    0        &        1     &   2  \\
\hline	
$^{216}$Th     &    V1        &   182.3 &  177.8  & 157.2  &  157.2  &   8.2   &   12.6   & 33.3  & 8.2    &    0        &        0    &   1    \\
\hline
\hline
\multicolumn{13}{|c|}{D1S, Symmetric path} \\
\hline	
 \end{tabular}
\caption{TKE and TXE evaluated at the exit point in the symmetric valley for the $^{230, 226, 222, 216}$Th isotopes. Coulomb, deformation energies and neutron multiplicities are 
added. Labels refer to the 3 scenarios. See text for explanations.
Calculations have been done with the D1S Gogny force. Energies are expressed in MeV.}
\label{TKED1S} 
\end{table*}

\begin{table*} \centering
\begin{tabular}{|c|c|c|c|c|c|c|c|c|c|c|c|c|}
\hline
 Nucleus    &   Valley     & ~~TKE$_{1}$~~ & ~~TKE$_{2}$~~ & ~~TKE$_{3}$~~ &~~ E$_{\rm coul}$~~                     &~~ TXE$_{1}$ ~~        & ~~TXE$_{2}$ ~~&~~TXE$_{3}$ ~~&~~ E$_{\rm def}$~~ & ~~ $ \nu_{1}$ ~~&~~ $ \nu_{2} $~~ &~~ $ \nu_{3} $~~    \\ 
\hline	
\hline
$^{230}$Th     &    V1     &  167.7                   &  153.2          &  151.3   &    151.3                   &  26.8                 &  41.3         &      43.2       &   26.8   &   1           & 2           &   2 \\
\hline
$^{226}$Th     &    V1     &  171.8                   &  161.0          &  152.4   &     152.4                  &  20.0                 &  30.7         &      39.4       &   20.0   &   1           & $\sim$2     &   2 \\
\hline	
               &{\bf  V2} &  {\bf 186.3}             &   {\bf 183.3}   &  {\bf 173.7}   &    {\bf 173.7}       &  {\bf 5.4}            &  {\bf 8.4}    &      {\bf 18.0} &    5.4  &   {\bf 0}     & {\bf 0}     &{\bf  1}\\
\hline
$^{222}$Th     &    V1     &  175.1                   &  166.8          &  156.1    &     156.1                 &  15.5                 & 23.9          &      34.5       &   15.5   &   0           &  1          &   2   \\
\hline
               &{\bf  V2} &  {\bf 187.5}             &  {\bf 185.7}    &  {\bf 171.8}   &     {\bf 171.8}      &  {\bf 3.2}            &  {\bf 4.9}    &      {\bf 18.9 }&    3.2  &   {\bf 0}     & {\bf 0}      &{\bf  1}\\
\hline	
$^{216}$Th     &    V1     &  180.0                   &  174.7          &   160.8  &      160.8                 &  9.8                  & 15.1          &      29.0       &    9.8  &   0           &   $\sim$1    &   1   \\
\hline	
               &{\bf  V2} &  {\bf 187.5}             &  {\bf 186.2}    &  {\bf 172.1}   &     {\bf 172.1}      &  {\bf 2.4}            & {\bf 3.7}     &      {\bf 17.8} &    2.4  &   {\bf 0}     &   {\bf 0}    &{\bf  1} \\
\hline	
\hline
\multicolumn{13}{|c|}{D1ST2a, Symmetric path} \\
\hline
 \end{tabular}
\caption{Same as Table \ref{TKED1S} but for the D1ST2a Gogny interaction.}
\label{TKED1ST2a} 
\end{table*}

We now focus on the number of neutrons that are emitted at the exit 
points in all the symmetric valleys. Scission ends up with two similar 
fragments and the neutron multiplicity has been calculated for each of 
them. First, one has made the assumption that all the available energy 
from the TXE transforms itself into neutron emission. Once all the 
possible neutrons are emitted, the rest of the TXE would be devoted to 
$\gamma$ emission, whose description is beyond the scope of this work. 
In order to account for $\gamma$ emission, GCM + particle number 
projections techniques \cite{RodPRL07, BenPRC08, YaoPRC10, BerPRC16} 
should be used on each of the fragments. Thus, the TXE writes:
\begin{align}
\mathrm{TXE}  &=\sum_{\rm i=1}^{2} \mathrm{E}^{(\rm i)}_\gamma + \nu^{(\rm i)} \langle\mathrm{E}^{(\rm i)}_{\rm n}\rangle + \sum_{\rm j=1}^{\nu^{(\rm i)}} 
\mathrm{S}_{\rm n}^{(\rm j)}  
\label{TXEng}
\end{align}
where $\mathrm{E}^{(\rm i)}_\gamma$ is the part of energy used to emit 
$\gamma$ in fragment $\rm i$, $\langle\mathrm{E}^{(\rm i)}_{\rm 
n}\rangle$ is the mean neutron kinetic energy, $\mathrm{S}_{\rm 
n}^{(\rm j)}$ are the successive one neutron separation energy of the 
fragment up to the post neutron emission fission product: 
$\mathrm{S}_{\rm n}^{(\rm j)}=\mathrm{S}_{\rm n}(\rm Z_{\rm frag},\rm 
N_{\rm frag}- \rm j+1)$. 
symmetric fission, the summation over the fragment's label $i$ in Eq. 
(\ref{TXEng}) can be replaced by an overall factor 2. Separation 
energies are presented in Table \ref{neutron}. A comparison of D1S and 
D1ST2a HFB calculations with experimental data leads to the conclusion 
that both interactions give a satisfactory agreement with experiment 
with a deviation of a few hundred keV.

\begin{table}
 \begin{tabular}{|l|c|c|c|c|c|}
 \hline
\rm Isotope & $^{115}$Rh & $^{114}$Rh & $^{113}$Rh & $^{112}$Rh & $^{111}$Rh \\
\hline
\rm S$_{\rm n}^{\rm D1S}$    & 6.224      & 4.918       &  7.139     & 5.174       & 7.654     \\
\rm S$_{\rm n}^{\rm D1ST2a}$ & 6.907      & 4.771       &  7.321     & 5.075       & 7.638     \\
\rm S$_{\rm n}^{\rm exp}$    & 6.590      & 5.020       &  7.110     & 5.500       & 7.547      \\
\hline
\hline
\rm Isotope & $^{110}$Rh & $^{109}$Rh & $^{108}$Rh & $^{107}$Rh      &  $^{106}$Rh\\
\hline
\rm S$_{\rm n}^{\rm D1S}$    & 5.606      & 8.194      &  5.480     & 8.168  &  6.084    \\
\rm S$_{\rm n}^{\rm D1ST2a}$ & 5.476      & 8.067      &  5.898     & 8.605  &  4.569    \\
\rm S$_{\rm n}^{\rm exp}$    & 5.900      & 8.039      &  6.239     & 8.573  &      \\
\hline
 \end{tabular}
\caption{Experimental S$_{\rm n}^{\rm exp}$ and theoretical S$_{\rm n}^{\rm D1S,D1ST2a}$ one-neutron separation energy S$_{\rm n}$ in Rhodium isotopes. 
Energies are expressed in MeV.}
\label{neutron} 
\end{table}

Mean neutron kinetic energies $\langle$ E$^{(\rm i)}_n$ $\rangle$ are 
displayed in Table \ref{TKEn}. These quantities are evaluated using the 
GEF model of Ref. \cite{SchNDS16} for a neutron incident energy at the 
barrier.

\begin{table}
 \begin{tabular}{|l|c|c|c|c|}
 \hline
\rm Isotope                  & $^{230}$Th & $^{226}$Th & $^{222}$Th & $^{216}$Th  \\
\hline
$\langle$ $\rm E^{(\rm i)}_{\rm n}$ $\rangle$   & 1.924      & 1.814      &  1.855     & 1.852        \\
\hline
 \end{tabular}
\caption{Mean neutron kinetic energy $\langle$ E$^{(\rm i)}_{\rm n}$ $\rangle$ extracted from GEF. Energies are expressed in MeV.}
\label{TKEn} 
\end{table}

The neutron multiplicity of each fragment
$\nu^{(\rm i)}$ is  extracted from Eq.~(\ref{TXEng}) as the biggest integer which satisfies:
\begin{equation}
\mathrm{TXE}  \ge \sum_{i=1}^{2} \nu^{(\rm i)} \langle\mathrm{E}^{(\rm i)}_{\rm n}\rangle + \sum_{j=1}^{\nu^{(\rm i)}} \mathrm{S}_{\rm n}^{(\rm j)}  
\end{equation}
Neutron multiplicities $\nu^{(\rm i)}$ for each fragment are reported 
in Tables~\ref{TKED1S} and ~\ref{TKED1ST2a} for all the three different 
scenarios. A tilde is used when less than $500$~keV are missing in the 
TXE to reach the next integer value. The $\nu^{(\rm i)}$ globally 
decrease with the mass number for both interactions. The valley V2 does 
not provide neutron emission, except for scenario 3.

The emergence of the second symmetric valley V2 leads to a bigger TKE 
than for the first valley V1 and thus a smaller TXE. A drop of the 
experimental $\langle\nu_{tot}\rangle$ for symmetric fission is 
expected when going to light Thorium isotopes.

As already mentioned for D1ST2a, the second valley V2 is not 
energetically favored in the $^{226}$Th and $^{222}$Th isotopes, 
contrary to the $^{216}$Th case. When comparing $^{222}$Th or 
$^{216}$Th ( valley V2 ) with $^{230}$Th (valley V1) the Coulomb energy 
E$_{\rm coul}$ is higher by $\simeq 20$~MeV in the lighter isotopes. 
This additional kinetic energy corresponding to a compact scission mode 
results in a drop of the TXE which leads to a loss of 1 neutron per 
fragment for scenario 1 and 3, and 2 neutrons per fragment for scenario 
2. It is in agreement with the loss of $2-2.5$ neutrons on the total 
multiplicity which has been measured by the SOFIA group 
\cite{SOFIA1,SOFIA2}.

\section{Conclusion and perspectives}\label{conclusion}

In this work, the effect of the tensor term on fission paths has been 
studied for the first time. In that context, we have investigated the 
asymmetric to symmetric fission transition in the light thorium 
isotopes which experimentally hints to the existence of a new, compact 
and symmetric, fission mode. We have used a static calculation based on 
an axial HFB approach breaking reflection symmetry and introducing 
constraints on multipole moments and particle numbers. Both, the D1S 
and the D1ST2a (D1S plus a perturbative finite range tensor) Gogny 
interactions have been used.

We have shown that, depending on the isotope, the tensor term can 
change the barrier height in a non negligible way. In particular, it is 
able to re-equilibrate the second hump height between the asymmetric 
and symmetric path. Indeed, in the $^{222}$Th isotope, this difference 
with the D1S interaction has been found equal to $\simeq 4$~MeV whereas 
it is reduced to $1.5$~MeV with the D1ST2a interaction. Thus, the tensor 
interaction renders the pure symmetric path more probable. In the 
$^{216}$Th isotope, this difference disappears.

Another striking feature is the appearance of a second valley in the 
\{Q$_{20}$,Q$_{40}$\} collective variables in the presence of the tensor term. 
Its existence and its deformation characteristics (much smaller values of Q$_{20}$ and 
Q$_{40}$ than the ones of the standard valley which leads to the known 
well-elongated symmetric fission) are interpreted as the theoretical 
proof of the experimentally observed symmetric compact fission mode.
It is the most remarkable result of this analysis.

The present study does not consider the dynamical aspects of fission 
and therefore cannot predict fission fragment mass distributions. In 
order to describe the population of the various valleys, it will be 
very interesting to perform a 3-dimensional dynamical calculations 
including Q$_{20}$, Q$_{30}$ and Q$_{40}$ as collective variables. 
Moreover, it would be crucial to perform systematic calculations in 
order to localize the possible areas where the tensor term is expected 
to play an important role for fission process. Finally, a full refit of 
the Gogny interaction including a finite range tensor term would be of 
prime interest.

\acknowledgments{}
R.N.B. and N.P. would like to thank A. Chatillon and J. Taieb for 
valuable discussions on SOFIA data in Thorium isotopes. They would 
like to thank also J.-F. Berger for his valuable advice 
and N. Dubray for his comments on the present work. 
This work has been supported by the Spanish Ministerio de Ciencia, Innovaci\'on y Universidades and the 
European regional development fund (FEDER), grants Nos FIS2015-63770, 
FPA2015-65929, PGC2018-094583-B-I00 (LMR), FPA2015-67694-P and PFPA2015-67694-P.


\end{document}